\documentclass{aa}  

\usepackage{txfonts}
\usepackage{graphicx}

\usepackage[breaklinks,
            pdftitle={SVEEEETIES},
            pdfsubject={Extrasolar planets, atmospheric spectroscopy, remote sensing},
            pdfkeywords={Radiative transfer;  Planets and satellites: atmospheres;  Techniques: spectroscopic; ­Infrared: planetary systems; Methods: data analysis},
            pdfauthor={F. Schreier et al.}]{hyperref}

\newcommand{\qeq}[1]  {Eq.~(\ref{#1})}
\newcommand{\qufig}[1] {Fig.~\ref{#1}}
\newcommand{\T} {{\mathcal T}}
\newcommand{\D}{\mathrm{d}}
\newcommand{\mtx}[1] {\mathsf{#1}}             
\newcommand{\mtxT}[1] {\mathsf{#1}^\text{T}}   
\newcommand{\cm}  {\ensuremath{\rm\, cm^{-1}}}
\newcommand{\mue} {\ensuremath{\rm\, \mu m}}

\begin{document}

\title{\LARGE\bf SVEEEETIES: \\ \Large Singular vector expansion to estimate Earth-like exoplanet temperatures from infrared emission spectra}

\author{F. Schreier\inst{\ref{imf}}
 \and   S. St\"adt \inst{\ref{imf}}
 \and   F. Wunderlich\inst{\ref{tub}}
 \and   M. Godolt\inst{\ref{tub}}
 \and   J. L. Grenfell\inst{\ref{ipf}}
}

\institute{DLR --- Deutsches Zentrum f\"ur Luft- und Raumfahrt,
           Institut f\"ur Methodik der Fernerkundung,
           82234 Oberpfaffenhofen, Germany \email{franz.schreier@dlr.de}\label{imf}
 \and
           TUB --- Technische Universit\"at Berlin,
           Zentrum f\"ur Astronomie und Astrophysik,
           Hardenbergstr.\ 36, 10623 Berlin, Germany\label{tub}
 \and
           DLR --- Deutsches Zentrum f\"ur Luft- und Raumfahrt,
           Institut f\"ur Planetenforschung,
           Rutherfordstr.\ 2, 12489 Berlin, Germany\label{ipf}
}%
\date{Received: 14 August 2019 / Accepted: 08 December 2019}

\abstract
{Detailed characterizations of exoplanets are clearly moving to the forefront of planetary science.
Temperature is a key marker for understanding atmospheric physics and chemistry.}
{We aim to retrieve temperatures of \element{N_2}-\element{O_2} dominated atmospheres from secondary eclipse spectroscopic observations of the thermal emission of Earth-like exoplanets orbiting G-, K-, and M-stars, using large-aperture future space telescopes.}
{A line-by-line radiative transfer code was used to generate synthetic thermal infrared (TIR) observations.
The atmospheric temperature is approximated by an expansion with the base vectors defined by a singular value decomposition of a matrix comprising representative profiles.
A nonlinear least squares fitting was used to estimate the unknown expansion coefficients.} 
{Analysis of the $4.3\rm\,\mu m$ and $15\rm\,\mu m$ \element{CO_2} bands in the TIR spectra permits the inference of temperatures even for low signal-to-noise ratios (S/N) of 5 at medium resolution.
Deviations from the true temperature in the upper troposphere and lower-to-mid stratosphere are usually in the range of a few Kelvin, with larger deviations in the upper atmosphere and, less often, in the lower troposphere.
Although the performance of the two bands is equivalent in most cases, the longwave TIR is more favorable than the shortwave due to increased star-planet contrast.
A high spectral resolution, as provided by the James Webb Space Telescope (JWST) instruments, is important for retaining sensitivity to the upper atmosphere.
Furthermore, the selection of an appropriate set of base functions is also key.}
{Temperature in the mid-atmosphere, relevant for understanding habitability, can be suitably characterized by infrared emission spectroscopy with a resolution of at least
1000 (ideally $\approx$2500).
Obtaining the necessary S/N will be challenging even for JWST, however, it could be feasible with future space missions, such as the Origins Space Telescope or the Large Interferometer for Exoplanets.
In the meantime, a least squares fitting with an appropriate set of base functions is also applicable for other classes of planets.}


\titlerunning{Singular vector expansion for exoplanet temperature retrievals}
\authorrunning{F.\ Schreier et al.}

\maketitle


\section{Introduction}
\label{sec:intro}

About a quarter century after the first detection of an extrasolar planet \citep{Mayor95}, exoplanet science is placing a major focus on the characterization of these distant worlds.
There is a possible bias among the sample of about 4100 exoplanets detected so far (\url{http://exoplanet.eu/}) due to detection limits.
The search for these objects is still important and several space missions, such as TESS \citep{Ricker15} and PLATO \citep{Rauer14etal}, as well as ground-based programs, such as CARMENES \citep{Quirrenbach16}
and HARPS \citep{Mayor13}, are currently operational or in development.

In view of the large diversity in the planets discovered so far --- way beyond that of our solar system --- a detailed investigation is desirable.
Accordingly, spectroscopic observations by missions such as Hubble and Spitzer have been used to infer the atmospheric properties of exoplanets (mostly hot Jupiters) despite rather limited data quality
(essentially due to noise and resolution), and dedicated missions, such as ARIEL \citep[Atmospheric Remote-sensing Exoplanet Large-survey][]{Tinetti18etal} or CHEOPS \citep[Characterising Exoplanet Satellite][]{Benz18}, are underway.
Moreover, the James Webb Space Telescope (JWST), equipped with several infrared instruments, will greatly advance exoplanet science.
Likewise, ground-based, high-resolution observations have been analyzed successfully \citep[e.g.,][]{deKok14,Snellen14,Birkby18} and the forthcoming Extremely Large Telescope (ELT) \citep{Rodler18} will further improve the capabilities of atmospheric retrievals.

A fundamental challenge of exoplanet characterization is that bulk, surface, and atmospheric conditions must be inferred from remote observations.
Inverse problems \citep{Craig86} are challenging in general because of their ill-posed nature; for exoplanets, this is even more demanding because of the weak signal and lack of a priori knowledge.
Nevertheless, given the increasing quality of spectroscopic observations, a variety of retrieval codes have been developed, including\ CHIMERA \citep{Line12}, $\tau$-REX \citep{Waldmann15t,Waldmann15e},
AURA, HYDRA, and POSEIDON \citep{Gandhi17,Pinhas18,MacDonald17}, HELIOS-RETRIEVAL \citep{Lavie17}, as well as petitRadTrans \citep{Molliere19}, which exploit different radiative transfer modeling approaches
(line-by-line, k-distribution, scattering, etc.) and inversion techniques (optimal estimation, Monte Carlo Markov chain, etc.).
Moreover, codes originally developed for Earth or other solar system planets have been applied successfully for exoplanets \citep[e.g.,\ NEMESIS by][]{Irwin08}.

Temperature is one of the most fundamental parameters of atmospheric physics and chemistry.
Remote sensing of Earth's atmospheric temperature is done operationally using microwave and thermal infrared (TIR) sensors aboard a fleet of satellites \citep[e.g.,][]{Menzel18}
and is well-established for the characterization of solar system planets \citep[e.g.,][]{Hanel03}.
In the recent past, atmospheric spectroscopy and inversion techniques have also been used to infer exoplanet temperatures; for more, see the reviews by \citet{Madhusudhan18,Deming19}.

Temperature is a function of altitude (or pressure) but for the solution of the inverse problem, a discretization is required.
For exoplanet retrievals, a parameterized representation with a few free parameters is quite common \citep[e.g.,][]{Madhusudhan09}, whereas for Earth, level-by-level (or layer-by-layer) approaches are widespread.
\citet{Line13} emphasize that the second approach is invalid for most current exoplanet spectra due to their limited information content.
\citet{Barstow13g,Barstow13e} use optimum estimation \citep{Irwin08} to assess the retrievability of the atmosphere of the super-Earth GJ1214b and hot Jupiters, respectively;
they compared three different parameterizations and concluded that the proposed EChO space telescope \citep{Tinetti15} should have been able to characterize the temperature of gas giants with about 200\,K precision.
The $\tau$-REX \citep{Waldmann15e} employs a two-stage hybrid approach combining the parametric and layer-by-layer representation.

As an alternative to parameterized temperature representations or level-by-level discretization, an expansion of the temperature profile using an appropriate set of basis function can be used.
Singular value decomposition (SVD) and principal component analysis (PCA) have been established as powerful tools for solving inverse problems \citep[e.g.,][]{Jarchow98,Lindstrot12,Waldmann15t,Damiano19,Fan19}.
B-spline approximations of the unknown profile are another widely used approach for solving inverse problems \citep[e.g.,][]{OSullivan85,Doicu05b}.

Here we use the first left singular vectors resulting from an SVD of a matrix comprised of representative temperature profiles as base vectors and consider the expansion coefficients as
unknowns of a nonlinear least squares fit.
For a thorough test of this approach, we first considered a set of Earth climatological data with microwave and TIR synthetic spectra before applying our solver to synthetic TIR observations of Earth-like exoplanets orbiting M-dwarfs and G- and K-stars.

The organization of this paper is as follows:
the next section describes our methodology (forward modeling, least squares, and SVD) along with the data.
After presenting some further preliminaries, we demonstrate the retrievability of temperature for Earth and Earth-like exoplanets in Section \ref{sec:results}.
We continue with a discussion in Section \ref{sec:disc} and provide our conclusions in Section \ref{sec:concl}.
Some supplementary plots are provided in an online appendix.


\section{Theory}
\label{sec:theory}

\subsection{Forward model --- infrared radiative transfer}
\label{ssec:irrt}

The intensity (radiance) $I$ at wavenumber $\nu$ received by a downlooking observer can be described by the integral form of the equation of radiative transfer (neglecting scattering
and assuming local thermodynamic equilibrium) \cite[]{Liou80,Goody89,Hanel03,Zdunkowski07}
\begin{align} \label{schwarzschild}
 I(\nu) ~&=~  I_\text{surf}(\nu) ~ \T(\nu,0) ~+~ \int_0^\infty B(\nu,T(\tau')) ~ \exp{\bigl(-\tau'(\nu)\bigr)} \, \D \tau' \\
\label{schwarzschildWgtFct}
        ~&=~  I_\text{surf}(\nu) ~ \T(\nu,0) ~-~ \int_0^\infty B(\nu,T(z')) ~ {\partial \T(\nu,z') \over \partial z'} \, \D z' ~,
\end{align}
where $B$ is the Planck function at temperature $T$, and $I_\text{surf}$ is a background contribution, that is,\ surface thermal emission at bottom of the atmosphere (BoA).
The monochromatic transmission $\T$ and optical depth $\tau$ between observer and altitude $z$ are given according to Beer's law by
\begin{align} \label{beer}
 \T (\nu,z) ~&=~ \exp\bigl(-\tau(\nu,z)\bigr) \\ \notag
             &=~ \exp \left[ - \int_z^\infty \sum_m k_m\bigl(\nu,p(z'),T(z')\bigr) \: n_m(z') \; dz' \right] ~,
\end{align}
where $n_m$ is the density of molecule $m$ and $k_m$ the pressure and temperature-dependent absorption cross section obtained by summing over the contributions from many lines, $k_m = \sum_l k_m^{(l)}$.
The finite spectral resolution of the instrument is taken into account by a convolution of the monochromatic radiance \eqref{schwarzschild} (or transmission \eqref{beer}) with a spectral response function (SRF).

The radiance is a nonlinear function of the quantities of interest for atmospheric retrievals, hence, any estimate of these atmospheric state parameters has to rely on iterative techniques.
For the Earth's temperature, sounding observation of the thermal radiation in spectral regions with dominant absorption by uniformly mixed gases such as carbon dioxide or oxygen is ideal,
that is, of the rotation band of \element{O_2} in the microwave around $2\rm\,cm^{-1}$ (60\,GHz) or the \element{CO_2} bands in the TIR around $700\rm\,cm^{-1}$ (or $15\mue$, $\nu_2$ band) and
$2400\rm\,cm^{-1}$ ($4.3\mue$, $\nu_3$ band).

In this context, the partial derivative in \eqref{schwarzschildWgtFct}, known as the weighting function $K(\nu,z) \equiv \partial \T(\nu,z) / \partial z$, is significant and it characterizes the contribution of an altitude layer to the outgoing radiation.
In particular, in these bands there is a one-to-one correspondence between altitudes and wavenumbers, indicating that in the band center radiation is predominantly coming from upper atmospheric layers,
whereas, in the band wings with smaller attenuation, radiation from the lower layers can penetrate through the atmosphere towards an observer located far away.

\begin{figure*}
 \centering\includegraphics[width=\textwidth]{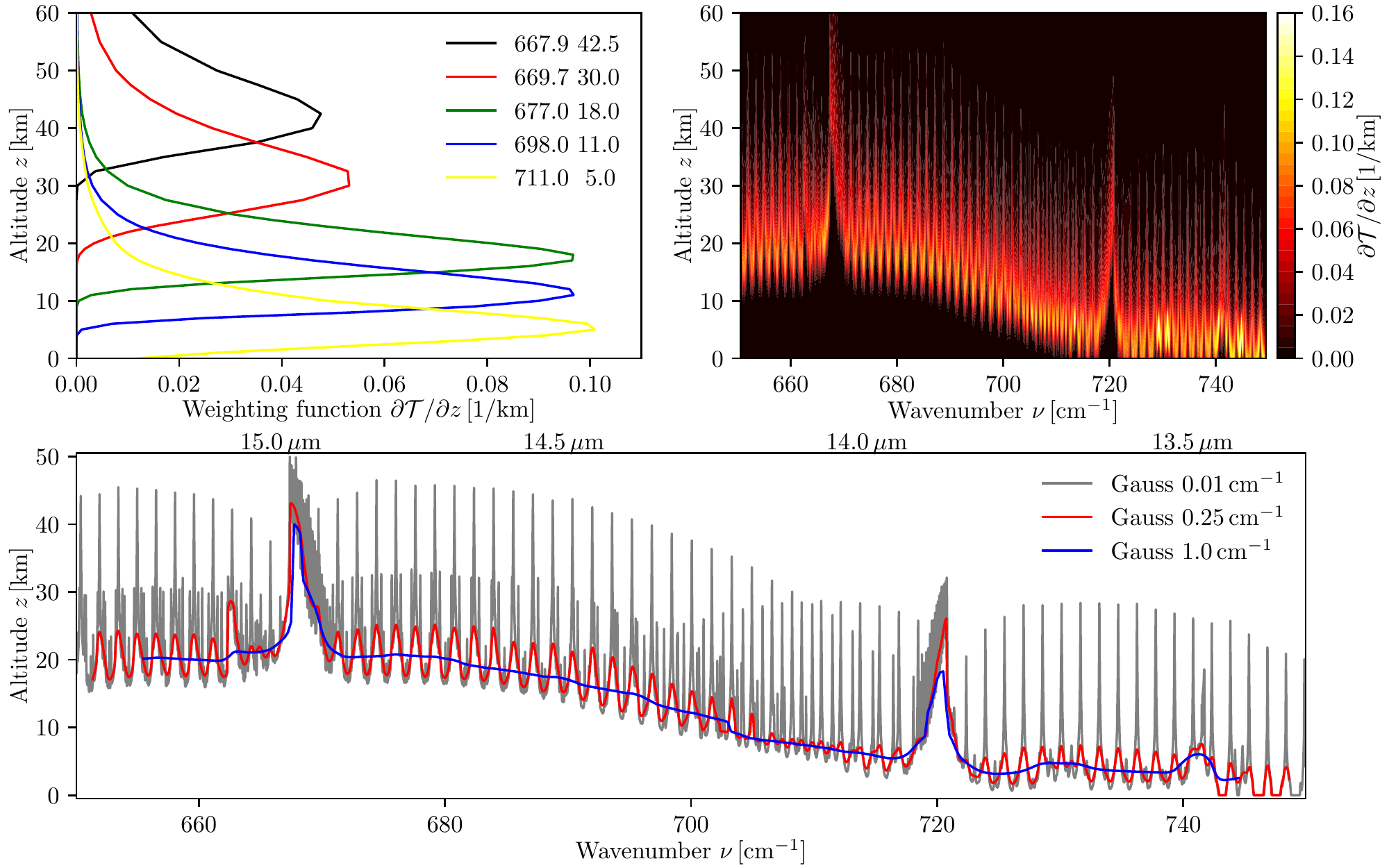}
 \caption{Weighting functions in the thermal infrared (\element{CO_2} $\nu_2$ band at $15\rm\,\mu m$, Earth's SAS).
          The upper panels show the weighting functions convolved with a Gaussian response function of half width $\Gamma=0.1\rm\,cm^{-1}$.
          Lower panel shows the peak altitude as a function of wavenumber for different Gaussian response functions.
          The $\Gamma=0.25\rm\,cm^{-1}$ Gaussian corresponds to a resolution $R \approx 2800$.
          Numbers in the top-left legend indicate the wavenumber [cm$^{-1}$] and the corresponding peak altitude [km].}
 \label{wgtFct_tir2}
\end{figure*}

\begin{figure*}
 \centering\includegraphics[width=\textwidth]{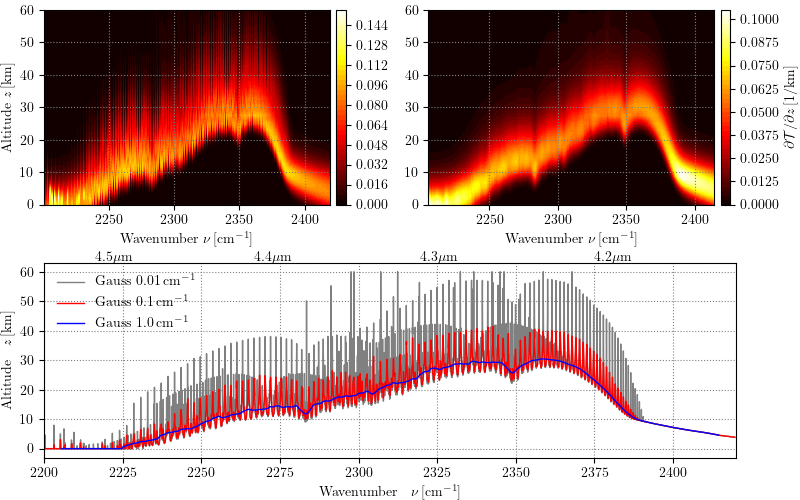}
 \caption{Weighting functions in the thermal infrared (\element{CO_2} $\nu_3$ band at $4.3\rm\,\mu m$, Earth's SAS).
           Plots on the top show the weighting functions convolved with a Gaussian response function of half width $\Gamma=0.1\rm\,cm^{-1}$ (left) and $\Gamma=1.0\rm\,cm^{-1}$ (right).
          Bottom plot shows the peak altitude of the weighting functions. The $\Gamma=1.0\rm\,cm^{-1}$ Gaussian corresponds to a resolution $R \approx 2300$.}
 \label{wgtFct_tir3}
\end{figure*}

Figures \ref{wgtFct_tir2} and \ref{wgtFct_tir3} show weighting functions in the TIR computed for Earth's subarctic summer atmosphere (SAS, see subsection \ref{ssec:data} below).
Clearly, the altitudes sensitive to the retrieval are linked to the spectral resolution.
Increased (decreased) \element{CO_2} concentration will lead to stronger (weaker) absorption and hence will shift the height of the peaks upwards (downwards).
Weighting functions of Earth-like planets around F, G, K, and M stars are rather similar, as discussed in \citet{Vasquez13m} (see also \qufig{proxCenWgtFct}).
For an analytical example assuming a single Lorentzian absorption line in the microwave, see \citet[][subsection 2.3.3.]{Doicu10}.

\subsection{Inversion methodology --- linearization and discretization}
\label{ssec:inv}

Assuming that the weighting function $K(\nu,z)$ is independent of temperature and neglecting the surface emission transmitted through the atmosphere, the Schwarzschild equation can be seen as a Fredholm integral equation of the first kind \citep{Craig86}:\ 
\begin{equation} \label{fredholm}
 y(\nu) ~=~ \int K(\nu,z) \, f(z) \: \D z ~.
\end{equation}
Here $y$ on the left corresponds to the observation (i.e.,\ the radiance spectrum as a function of wavenumber) and $f(z)$ in the integral corresponds to the atmospheric profile to be retrieved,
for example, temperature $T(z)$ (the Planck function $B$ increases monotonically with temperature).
In general, a Fredholm integral equation can be obtained by linearization of the Schwarzschild equation \eqref{schwarzschild} around the initial (or current) guess of the unknown profile
(temperature $T_\text{c}$ etc.). Then $y(\nu)$ corresponds to the radiance residual $I_\text{obs}(\nu) - I_\text{mod}(\nu,T_\text{c})$ and $f(z)$ corresponds to the change in the temperature profile.

To solve \qeq{fredholm} for the unknown profile $f(z),$ discretization is mandatory.
Since there is only a finite set of measurements available, the function $y(\nu)$ is essentially a measurement vector $\vec{y}$ with $m$ elements $I_\text{obs}(\nu_1),I_\text{obs}(\nu_2), \dots, I_\text{obs}(\nu_m)$.
For the profile function $f(z),$ there are essentially three approaches:
since the integral has to be evaluated by means of quadrature, that is, $\int K(\nu,z) f(z) \, \D z \approx \sum_{j=1}^n w_j K(\nu,z_j) f(z_j)$ with weights $w_j$, the function values at the quadrature
nodes can be considered as unknowns, leading to the ``state vector'' $\vec{x} = \bigl( f(z_1), f(z_2), \dots f(z_n) \bigr)$.
However, for a reasonable representation of the atmospheric profile some dozen nodes can be required, yet the solution of \eqref{fredholm} can often lead to an oscillatory profile because of
the ill-posed nature of the inverse problem.
To overcome the problem of zigzag solutions, regularization is mandatory, meaning that additional information, such as smoothness constraints, must be included \citep[e.g.,][]{Schimpf97}.

For exoplanet retrievals, parameterized temperature profiles are commonly used.
For the case of hot Jupiters,  \citet{Madhusudhan09} used three individual parameterisations for troposphere, stratosphere, and mesosphere with six free parameters,
whereas \citet{Line12} used an analytic temperature profile originally attributed to \citet{Guillot10} based on four parameters.
A similar analytic scheme \citep{Parmentier14} was used by \citet{Morley17gj} for Neptune-mass planets with six parameters
and by \citet{Benneke19} for a sub-Neptune with five parameters.
Parameterizations were also used for terrestrial planets by, for example, \citet{Paris13r} and \citet{Morley17jw}. 
\citet{Rocchetto16} found that retrievals using parameterized temperatures facilitate a better characterization of hot Jupiter atmospheres than an isothermal approximation.

As an alternative, the profile can be expanded in a set of basis functions $\phi_j, j=1, \dots, n$ according to:
\begin{equation} \label{profileExpansion}
 f(z) ~=~ \sum_{j=1}^n x_j \phi_j(z) ~,
\end{equation}
and the state vector is defined by the expansion coefficients $x_j$.
Utilizing an appropriate set of basis functions has already demonstrated some regularizing effect \citep[e.g.,][]{Doicu05b}.

\subsection{Linear least squares and singular value decomposition (SVD)}
\label{ssec:svd}

A linear least squares problem, such as that resulting from discretisation of the Fredholm equation \eqref{fredholm}, is defined by:
\begin{equation} \label{LinearLeastSquares}
\min_x \| \mtx{A} \vec{x} - \vec{y} \|^2~.
\end{equation}
Here $\mtx{A}$ is an $m\!\times\!n$ matrix with $A_{ij} \equiv w_j\,K(\nu_i,z_j)$, $\bf y$ is an $m$-vector of measurements (e.g.,\ spectrum $y_i \equiv I_\text{obs}(\nu_i)$ for $i=1,\dots, m$), and $\vec{x}$ is the unknown $n$-vector representing the sought profile $f(z)$ (e.g.,\ temperature $T(z)$).
Note that the better the discretisation (increasing $n$), the more similar become the rows of the matrix $\mtx{A}$, hence, the more ill-conditioned the linear system.

The SVD of an $m\times n$ matrix $\mtx{A}$ is given by the factorization \citep[e.g.,][section 6.8]{Kahaner89}
\begin{equation} \label{SVD}
 \mtx{A} ~=~ \mtx{U} \, \mtx{S} \, \mtxT{V}
         ~=~ \sum_{i=1}^r \sigma_i \, \vec{u}_i \, \vec{v}_i^\text{T},
\end{equation}
where $\mtx{U}$ and $\mtx{V}$ are unitary matrices with columns comprising the singular vectors $\vec{u}_i$ and $\vec{v}_j$, $i=1,\dots,m$ and  $j=1,\dots,n$, respectively,
and $\mtx{S}$ is an $m\times n$ diagonal matrix with $r\le\min(m,n)$ positive singular values $\sigma_i$ ordered by size, $\sigma_1 \ge \sigma_2 \ge \dots \ge \sigma_r > 0$. Here
$r$ is the rank of the matrix $\mtx{A}$, and $\vec{u}_i \, \vec{v}_i^\text{T}$ is the outer product of two vectors (the subscript indicates the ``transpose'').
The vectors $\vec{u}_i$ and $\vec{v}_j$ have an increasing number of sign changes with increasing $i,j$ and constitute a basis for $\mathbb{R}^m$ and $\mathbb{R}^n$, respectively.
The ratio of the first to the last singular value is denoted as the condition number of the matrix $\mtx{A}$, 
that is, $\kappa \equiv {\sigma_1 / \sigma_r}$, and characterizes the (ill-) conditioning of the linear system.

Truncating the sum in \eqref{SVD} at $k < r$ gives an approximation of the original matrix $\mtx{A}$ \citep[][section 6.8.2]{Kahaner89}.
The quality of this ``truncated SVD'' (TSVD) approximation depends on $r$ and the distribution of singular values $\sigma_i$.

Using the SVD \eqref{SVD}, the solution of \eqref{LinearLeastSquares} is obtained as expansion in the right singular vectors,
\begin{equation} \label{SVDsolution}
\vec{x} ~=~
\sum_{i=1}^r {\vec{u}_i^T \vec{y} \over \sigma_i} ~ \vec{v}_i ~.
\end{equation}
Ideally, the dot product in the numerator decays more rapidly than the singular values, so high frequency contributions are suppressed.
However, noise in the data vector $\vec{y}$ can prevent such decay and can lead to strongly oscillating solutions.
Hence, additional information is required, a process known as regularization.

\begin{figure*}
 \centering\includegraphics[width=\textwidth]{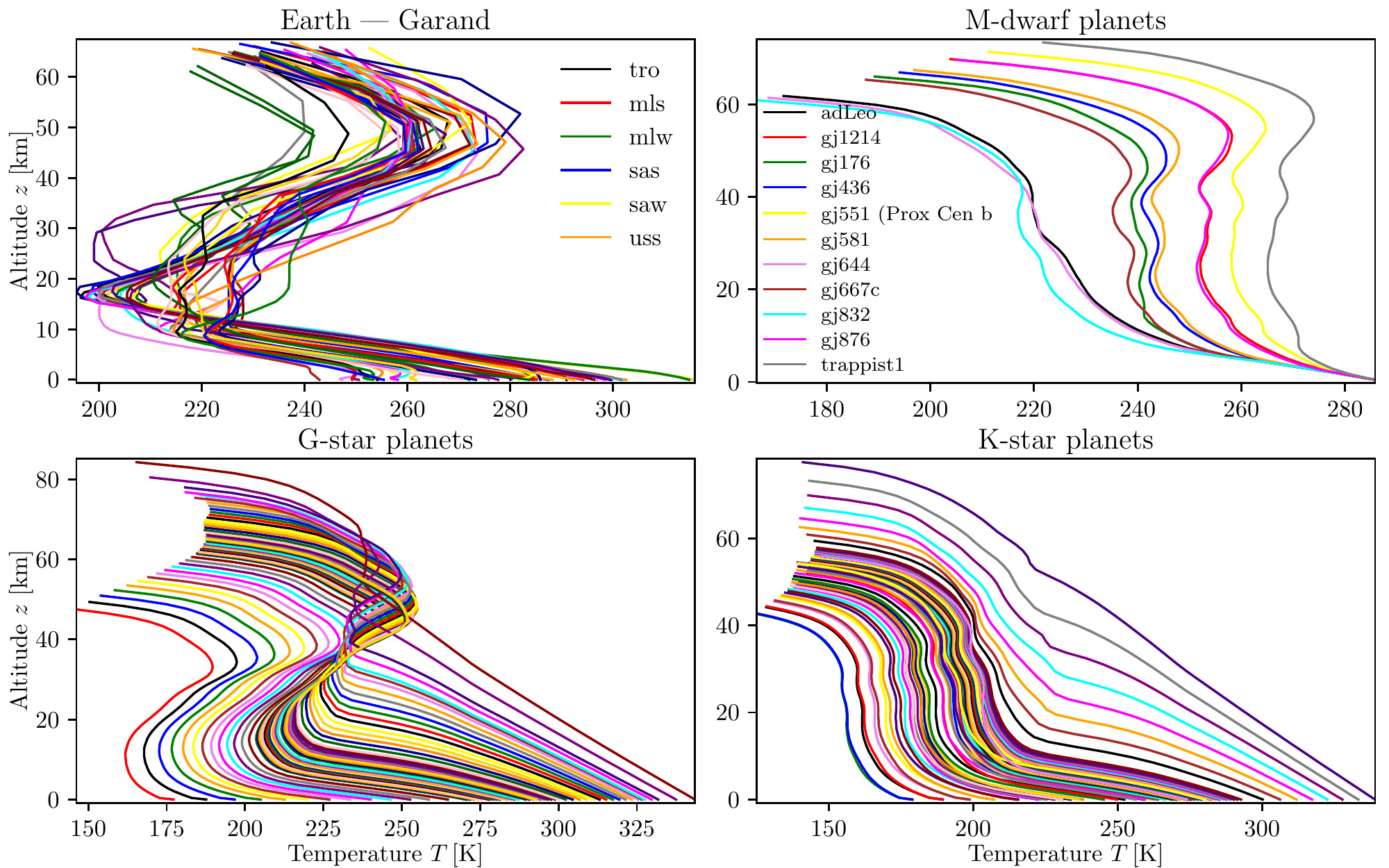}
 \caption{Temperature profiles of the 42 Earth--Garand atmospheres. Top left: the legend identifies the first six atmospheres:  tropical, midlatitude summer and winter, subarctic summer and winter, and US standard. M-dwarf planets are in the top right and G- and K-star planets are at the bottom.}
 \label{atmosTemp} 
\end{figure*}

\subsection{Input data}
\label{ssec:data}

For Earth, we use the set of 42 atmospheric profiles collected by \citet{Garand01}, comprising pressure, temperature, and molecular concentrations, which are representative of most
meteorological cloud-free situations, see \qufig{atmosTemp} (top left). (For the molecular concentration profiles see \qufig{garand_gases}.)
The first six atmospheres correspond to the AFGL (Air Force Geophysics Laboratory) atmospheres \citep{Anderson86}, atmospheres 7 -- 18, 19 -- 30, and 31 -- 42 are ranked by increasing mean temperature, water content, and total ozone, respectively.
The carbon dioxide volume mixing ratio (VMR) is constant at 360\,ppm for all atmospheres.
Originally, the data are given on 43 levels with a common pressure grid in the range $0.1$ to $1013.25\rm\,mb$ and slightly varying altitudes up to $61$ to $67\rm\,km$.

Exoplanet atmospheric scenarios are calculated with a 1D steady-state cloud-free climate model \citep[cf.][]{Paris15m}.
As input data for the Earth-like planets orbiting G- and K-type stars in their habitable zone we use the modeled temperature and concentration profiles from \citet{Godolt16}.
These temperature profiles are also depicted in \qufig{atmosTemp} (bottom; for an illustration of the complete dataset see \qufig{gkmData}).
Temperature and tropospheric water is calculated in the climate model by radiative transfer and convective adjustment under the assumption of a 0.22 surface albedo and an Earth-like composition.
These planets have a top of atmosphere (ToA) pressure of about $0.07\rm\,mb$.

For the M-dwarf Earth-like planets (henceforth called ``M-Earths'') we use data from \citet{Wunderlich19}; also see \qufig{atmosTemp} and  \qufig{gkmData}.
The atmospheric temperatures and composition are inferred from a 1D photochemistry model \citep[considering 55 species and 217 chemical reactions, cf.][]{Gebauer18e} coupled to the climate model.
The M-Earth data are defined on 64 levels with a ToA pressure of about $0.08\rm\,mb$.
For all M-Earths, methane VMR are almost constant with height (in contrast to Earth where the VMR decreases by about a factor 10 from BoA to ToA).
As discussed by \citet{Wunderlich19}, \element{CH_4} is strongly enhanced, with vertical column densities ranging from $6 \cdot 10^{21} \rm\,cm^{-2}$ for ``Earth'' placed
around GJ644 to $7 \cdot 10^{22} \rm\, cm^{-2}$ for Earth around Trappist\,1 --- compared to $3.5 \cdot 10^{19} \rm\,cm^{-2}$ or $1.6\rm\,ppm$ for the mean Garand atmosphere (modern Earth).
Nitrous oxide (\element{N_2O}) concentrations are also larger compared to Earth, especially in the upper atmosphere.
Water VMRs are almost identical at BoA and strongly decreasing in the upper atmosphere, with ToA VMRs varying by about two orders of magnitude.

All of these planetary atmospheres are assumed to be Earth-like, that is,\ 1 bar surface pressure, \element{N_2-O_2} dominated (78\% and 21\%, respectively), and 355\,ppm \element{CO_2} at BoA.
The tropospheric water vapour profile is taken from the climate model and is calculated using the relative humidity profile of Earth taken from \citet{Manabe67}.

These atmospheric data is used in two ways.
First, we combine the temperature profiles (regridded with piecewise cubic Hermite interpolation \citep{Kahaner89} to a common $z$ grid with $n_z$ equidistant levels) into a ``temperature matrix'' $\mtx{A}$ (e.g.,\ an $n_z \times 42$ matrix for the Garand data) and use the first $k$ columns of the left singular matrix $\mtx{U}$ resulting from the SVD \eqref{SVD} of this matrix as base vectors in the expansion \eqref{profileExpansion}.
Second, we use these atmospheric data to generate synthetic observations for all cases to test our retrieval procedure.
Unless otherwise noted, the profiles are discretized with an equidistant altitude grid of 31 points in $0 \le z \le 60\rm\,km$, that is,\ with $2\rm\,km$ altitude steps.

\subsection{Implementation}
\label{ssec:code}

For our forward model we use Py4CAtS --- Python for Computational ATmospheric Spectroscopy \citep[][available at \url{http://atmos.eoc.dlr.de/tools/Py4CAtS/}]{Schreier19p}, a Python re-implementation of the Generic Atmospheric Radiation Line-by-line Infrared Code (GARLIC) \citep{Schreier14}.
Despite the heavy use of Numerical Python --- NumPy \citep{vanderWalt11}, the runtime is considerably slower than GARLIC
\citep[which utilizes multi-threading and algorithmic differentiation, see][]{Schreier15} but Python was nevertheless preferred for this feasibility study because of its rapid prototyping and flexibility.

Py4CAtS computes (like GARLIC) molecular absorption cross sections assuming a (default) Voigt line shape \citep{Schreier18h}, where the wavenumber grid point spacing is 
adjusted automatically for each molecule, pressure and temperature to a fraction of the typical line width.
In contrast to most lbl models, our codes are level-oriented (i.e., without any\ Curtis-Godson approximation or the like for atmospheric layers).
The Schwarzschild equation \eqref{schwarzschild} is solved by numeric quadrature with either a ``linear-in-tau'' or an ''exponential-in-tau'' approximation for the Planck function.
Both observed and model spectra are convolved with a Gaussian spectral response function of constant width.

The synthetic measurement spectrum is generated by adding Gaussian noise (generated by the Numeric Python \texttt{numpy.random.randn} function) independent of wavenumber,
that is the signal-to-noise ratio (S/N) is assumed to be constant over the individual narrow wavenumber interval considered for the retrieval.
The inversion relies on the Scientific Python implementation \texttt{scipy.optimize.leastsq} of the MINPACK Levenberg-Marquardt nonlinear least squares solver \citep{More78}
that aims to minimize the difference between observed and model spectrum, essentially $\| \vec{y}-\vec{F}(\vec{x}) \|$ with $F$ representing the forward model, that is, the radiative transfer.


\section{Results}
\label{sec:results}

\subsection{Base vectors}
\label{ssec:base}

The proper selection of base vectors in the expansion \eqref{profileExpansion} is crucial for the success of the least squares fit.
The ``Garand atmospheres'' are representative for Earth, so the first singular vectors $\vec{u_1}, \dots, \vec{u}_k$ resulting from the SVD of the ``Garand temperature matrix'' should be suitable for the analysis of observations of Earth.
However, many exoplanets are not likely to show such a strong temperature inversion (cf.\ \qufig{atmosTemp}).
Accordingly, such exoplanet temperature profiles should be considered in the matrix to be decomposed according to \eqref{SVD} as well.

\begin{figure}
 \centering\includegraphics[width=\linewidth]{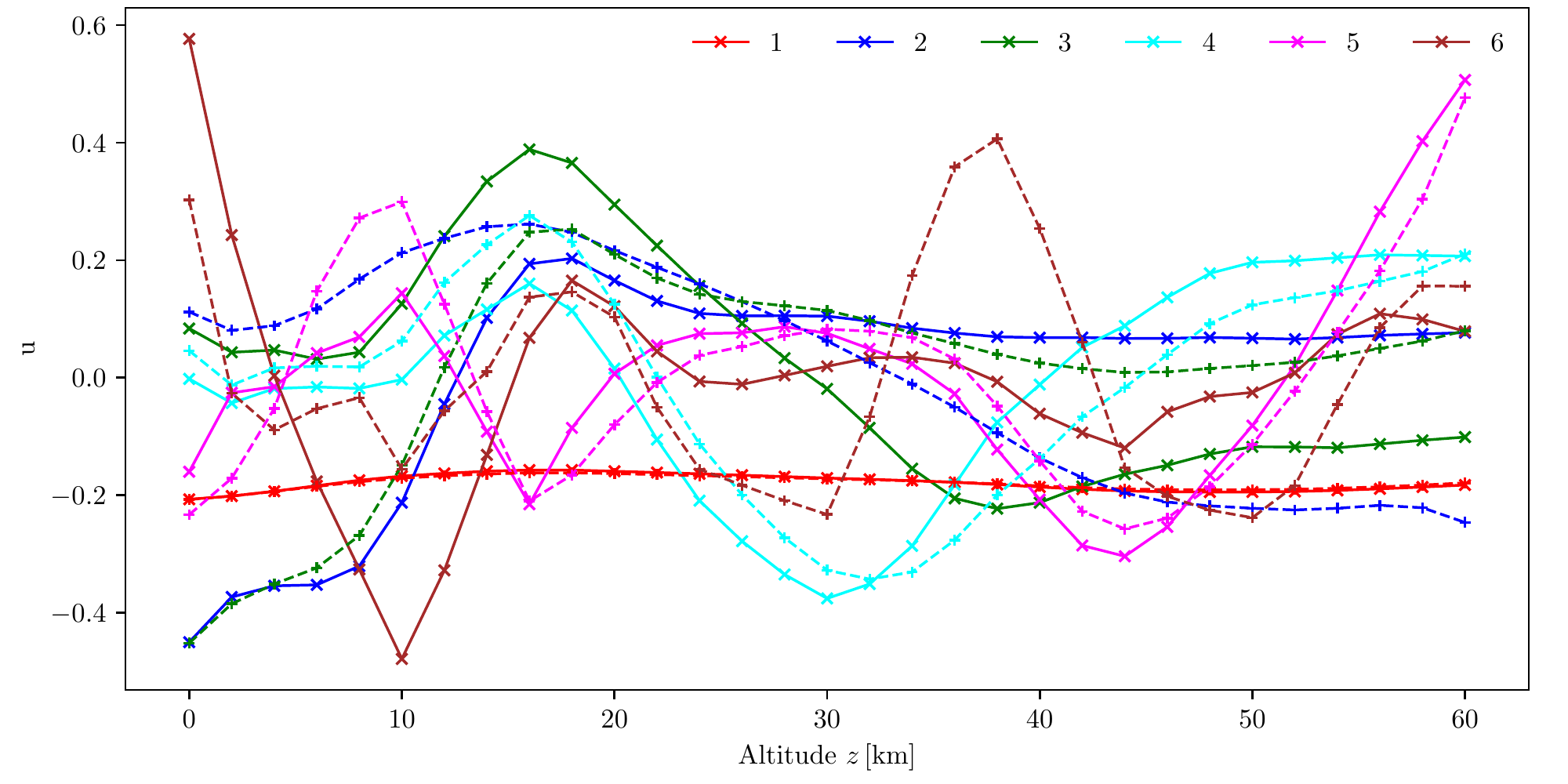}
 \caption{First six base vectors $\vec{u}$ resulting from the SVD of the matrix of Garand temperatures (solid lines marked with 'x') and Garand+MEarth profiles (dashed with '+' marker).}
 \label{baseVectors}
\end{figure}

Figure \ref{baseVectors} shows the left singular vectors $\vec{u_j}, ~ j=1,\dots,6$ resulting from the SVD of the ``Garand temperature matrix''
and the singular vectors resulting from the decomposition of the matrix comprising the 42 Garand and 11 M-Earth profiles.
The very first vector $\vec{u_1}$ of both decompositions are almost identical and the $\vec{u_4}$ vectors are quite similar.
Differences show up especially for the second and sixth vectors. 
We note that the first vector is approximately proportional to the mean temperature, that is,\ $\langle T \rangle \approx -1350 \, \vec{u_1}$ for Earth.

\subsection{Temperature reconstruction using linear least squares}
\label{ssec:linLS}

\begin{figure}
 \centering\includegraphics[width=\linewidth]{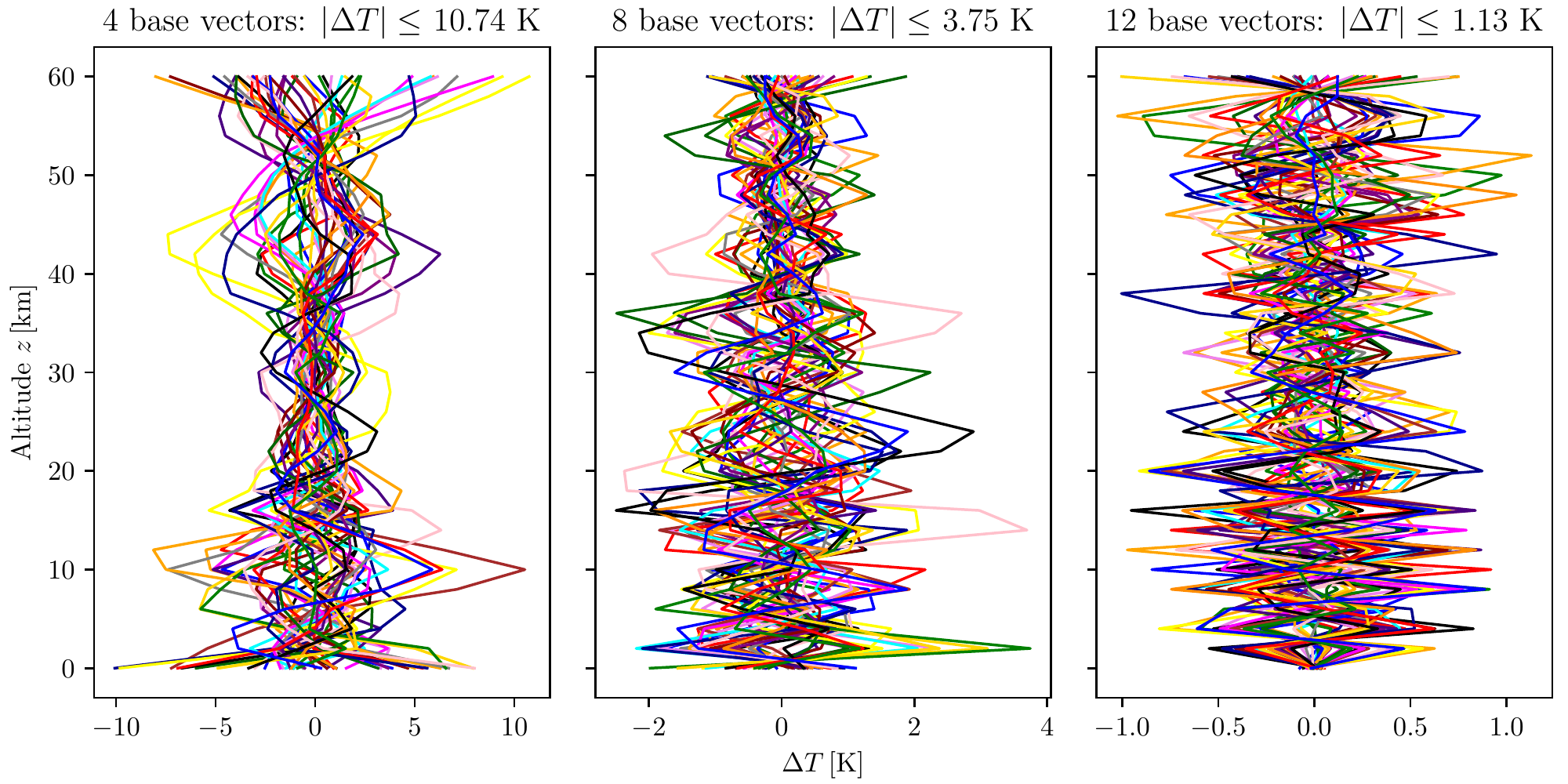}
 \caption{Errors of the 42 Garand temperature profiles reconstructed with four, eight, and twelve base vectors.}
 \label{garandUnSVD} 
\end{figure}

For a first assessment of the feasibility of temperature retrievals using the singular vectors as basis in the expansion \eqref{profileExpansion}, we study the quality of a linear least squares solution, that is,\ 
\begin{equation} \label{linLS}
\min_x \| \vec{T} - \sum_{j=1}^k x_j \vec{u}_j \|^2 ~.
\end{equation}
Figure \ref{garandUnSVD} shows the errors of the temperature profile vector $\vec{T}$ reconstructed with four to twelve base vectors.
With only four base vectors the errors can be as large as 10\,K, especially at BoA and ToA.
Doubling the number of base vectors reduces the errors by more than a factor of two, and 12 base vectors allow a reconstruction with errors smaller than 1\,K. 
The accuracy of meteorological temperature soundings using the AIRS \citep[Atmospheric Infrared Sounder,][]{Chahine06} and IASI \citep[Infrared Atmospheric Sounding
Interferometer,][]{Hilton12} instruments aboard the AQUA and MetOp satellites is also about one Kelvin \citep{Menzel18}. 

\begin{figure}
 \centering\includegraphics[width=\linewidth]{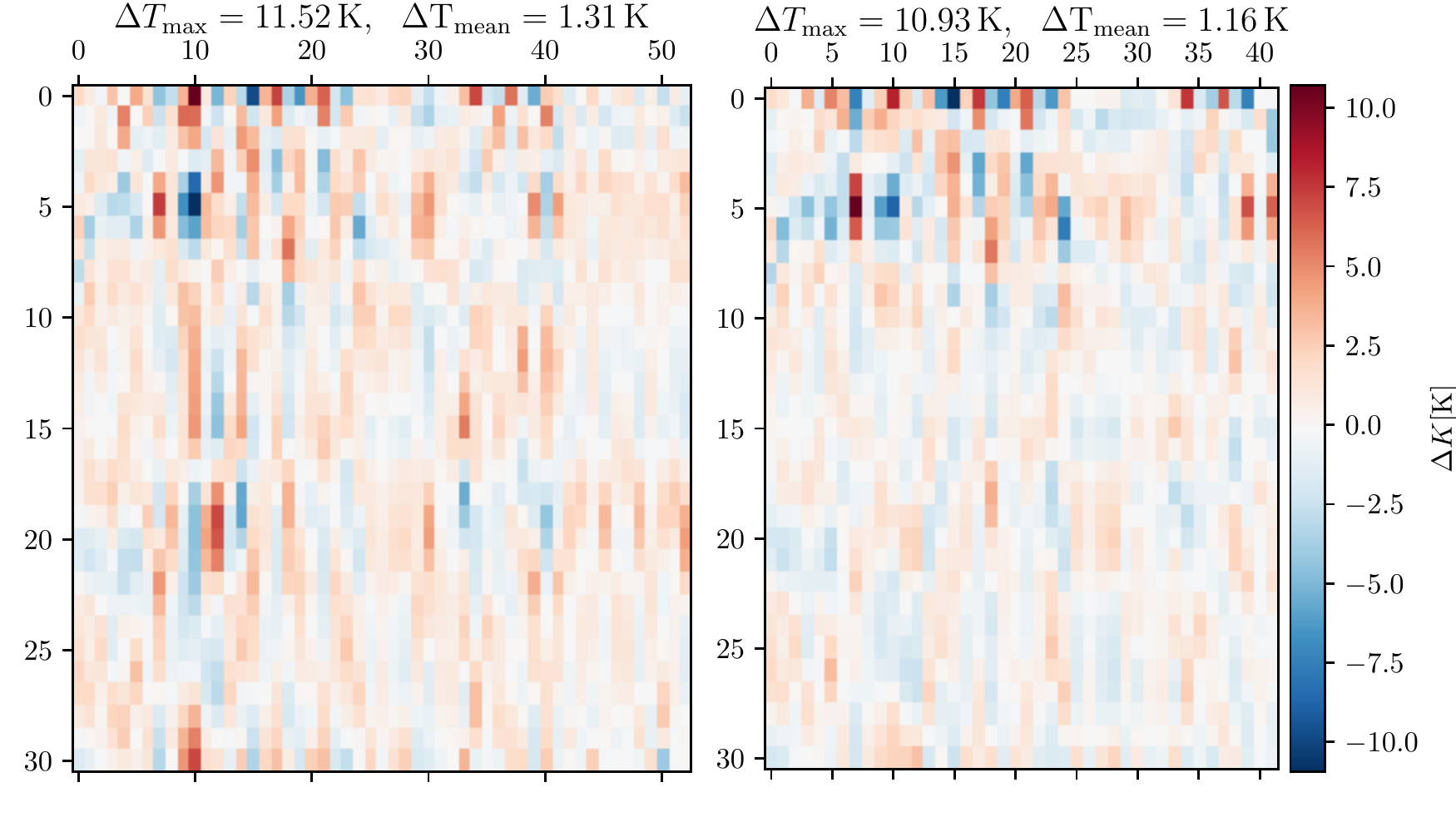}
 \caption{Errors of the truncated SVD reconstruction of the temperature profile matrix (defined on 31 equidistant levels $\le 60\rm\,km$, indices on the vertical axis) with five vectors.
Left:  SVD of 53*31 ``Garand+MEarth'' matrix with condition number 5.6e+04;
right: SVD of 42x31 ``Garand matrix'' with condition number 2.6+05.}
 \label{garandUnSVDmatrix} 
\end{figure}

Figure \ref{garandUnSVDmatrix} (right) shows the quality of the TSVD approximation applied to the temperature matrix comprising the set of Garand profiles (that is, a matrix of 31
rows and 42 columns).
This figure can be seen as an alternative visualization of the truncation error shown in \qufig{garandUnSVD}:
using the leading five terms in \eqref{SVD} (i.e., $k=5$) the error $|\Delta T|$ of the fitted temperature is mostly smaller than 5\,K, with some outliers as large as 10\,K in the upper stratosphere,
with $k=10$ the deviation is usually less than 2\,K (not shown).
Figure \ref{garandUnSVDmatrix} also shows the TSVD error using the combined Garand and M-Earths temperature matrix:
both the maximum and the mean error are slightly larger, and some larger deviations also show up in the lower stratosphere for some Earth profiles.
The M-Earth profiles are reproduced with errors less than five Kelvin.
However, the condition number (see Section \ref{ssec:svd}) is reduced by a factor of nearly five.

\begin{figure}
 \centering\includegraphics[width=\linewidth]{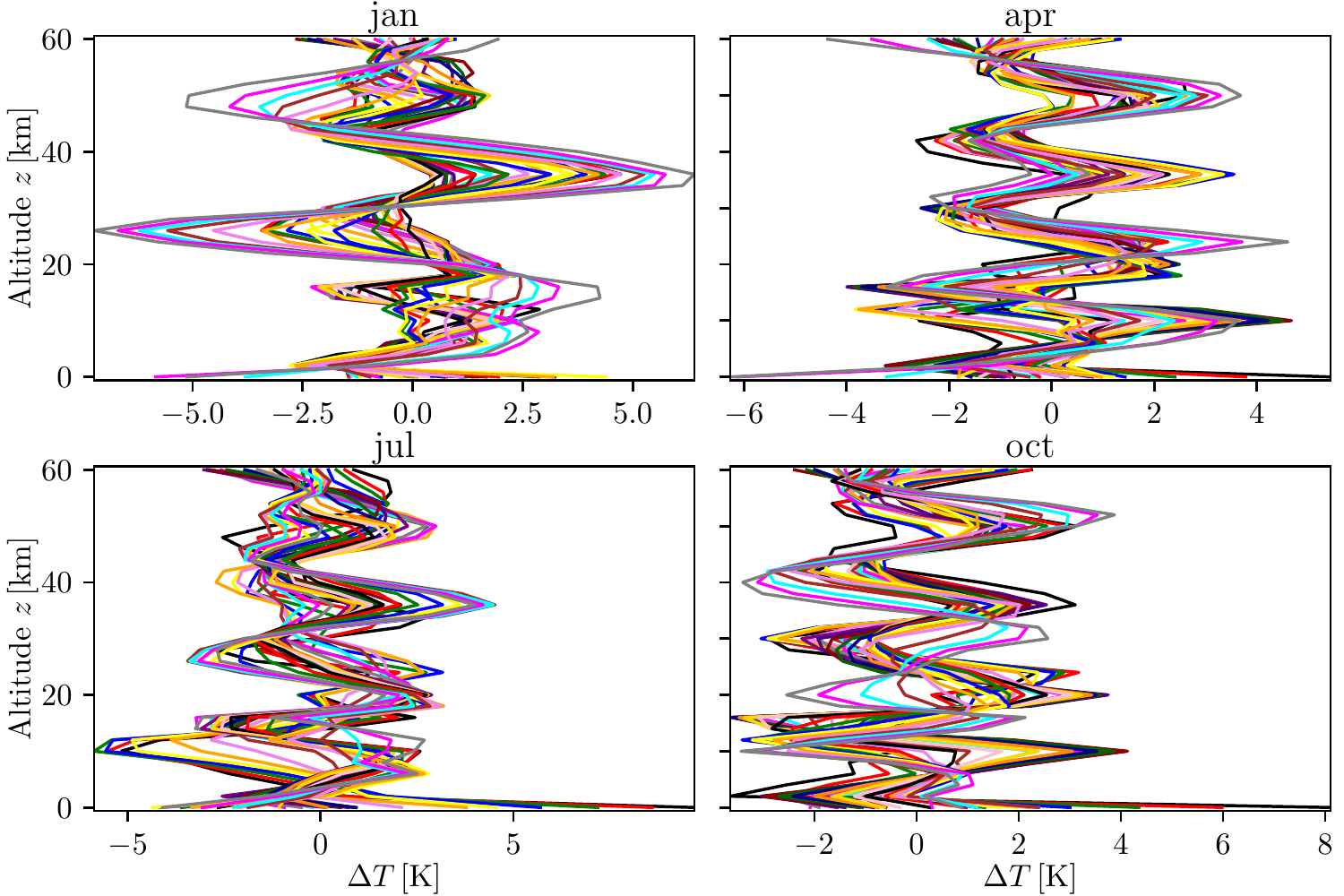}
 \caption{Errors of CIRA temperature profiles (January, April, July, and October) reconstructed with five base vectors.}
 \label{ciraSVD} 
\end{figure}

As a further test of the capabilities, we reconstructed the temperature profiles of the Committee on Space Research (COSPAR) International Reference Atmosphere \citep{Fleming90} available at \url{http://badc.nerc.ac.uk/data/cira/}.
CIRA provides monthly mean pressures and temperatures for the altitude range $0 \text{--}120\rm\, km$ extending almost from pole to pole in latitude steps of $5^\circ$.
Figure \ref{ciraSVD} shows that with five base vectors the temperatures can be fitted with errors less than $6\rm\,K$ except for somewhat larger deviations up to 10\,K in July and October. 

\begin{figure*}
 \centering\includegraphics[viewport=0 0 864 550, clip=true, width=\textwidth]{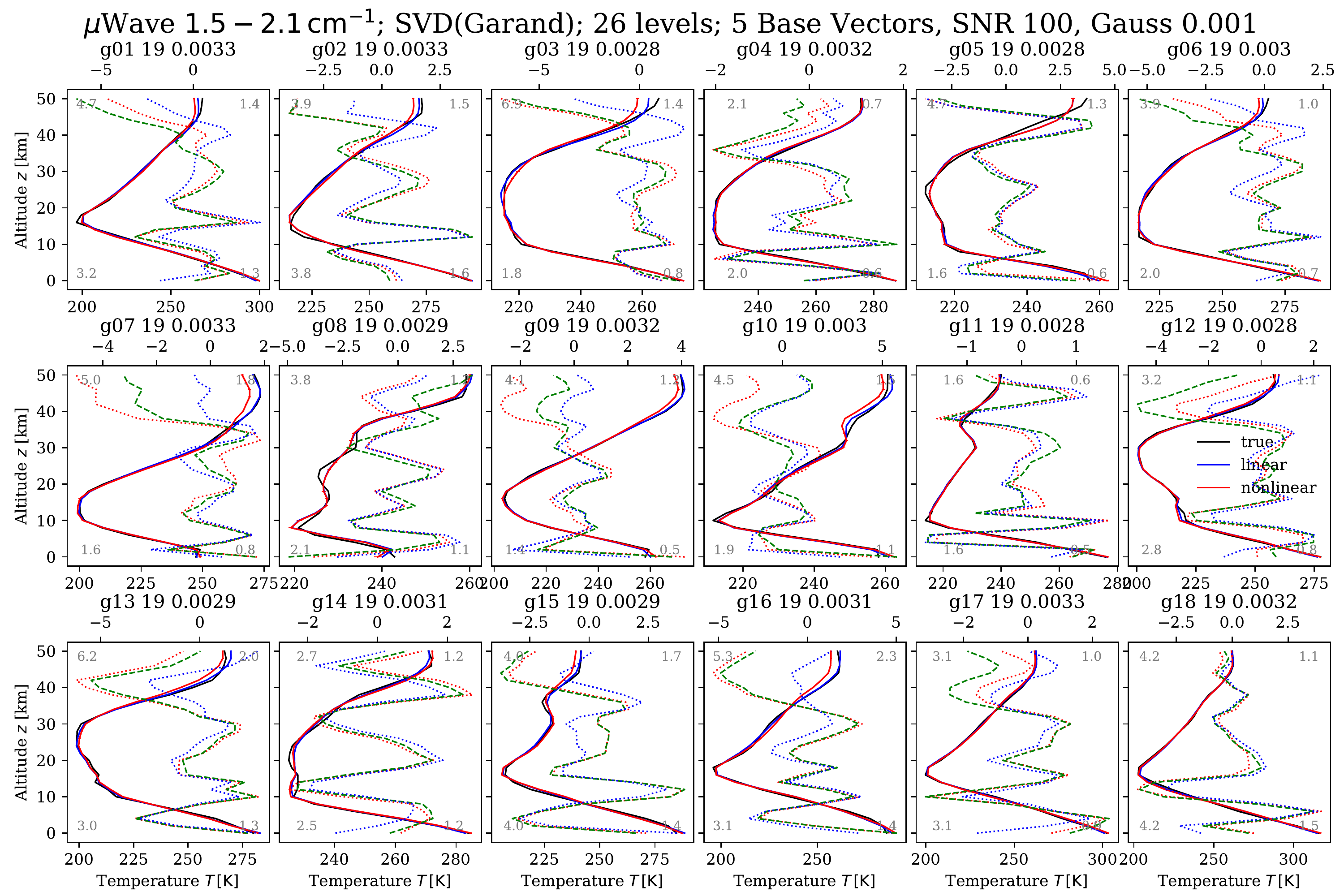}
 \caption{Microwave temperature least squares fits for the first 18 Garand atmospheres:
          $1.5 \text{--} 2.1\cm$ ($\approx 60\rm\,GHz$ or 5000\mue), S/N=100 and Gauss SRF with HWHM $\Gamma=0.001\cm$ ($R \approx 2000$).
          Black: true profile; blue: linear least squares reconstruction; red: nonlinear least squares with initial guess damped mean profile \eqref{initMean}.
          Temperature differences $\Delta T = T_\text{fit}-T_\text{true}$ are shown as dotted lines with axis at top.
          Dashed green: fit with initial guess straight line \eqref{initLine}.
          The title gives the atmosphere's ``name'', the number of iterations, and the squared norm of the radiance residuum.
          Numbers inside the plots indicate the maximum and mean temperature deviations (top) and corresponding UTLS deviations (bottom).}
 \label{garand_mWave_snr100} 
\end{figure*}

\subsection{Nonlinear least squares fits: preliminaries}
\label{ssec:init}

For the iterative solution of nonlinear optimization problems an initial guess is required, which is taken as a damped version of the mean temperature profile or as a straight line,
\begin{align}
 \label{initMean} T_\text{init}(z) ~&=~ 0.1 \cdot \bigl( T_\text{mean}(z) - T_\text{MoA} \bigr) + T_\text{MoA} \\
 \label{initLine} T_\text{init}(z) ~&=~ {z - z_\text{BoA} \over z_\text{ToA} - z_\text{BoA}} \cdot T_\text{ToA} + {z_\text{ToA} -z \over z_\text{ToA} - z_\text{BoA}} \cdot T_\text{BoA}
\end{align}
where $T_\text{mean}$ is the temperature resulting from averaging all 42 Garand atmospheres; the temperatures at bottom, mid, and top of atmosphere are chosen as
$T_\text{BoA}=260\rm\,K$, $T_\text{MoA}=250\rm\,K$, and $T_\text{ToA}=240\rm\,K$.

Pressure and concentrations of interfering gases (in the microwave and longwave (LW) TIR essentially water and ozone, in the shortwave (SW) TIR water and methane) are taken from the ``true'' atmosphere used to generate the synthetic measurement.

The errors of the fitted expansion coefficients $\vec x$ are estimated by the diagonal elements of the covariance matrix 
$ \| \vec y \| \sqrt{\left(\mtxT{J} \mtx{J}\right)^{-1} / (m-n)}$
, where $\mtx{J} = \partial \vec y / \partial \vec x$ and $\vec y$ correspond to the difference of observed and modeled radiance $I$.
The temperature error is then obtained by error propagation using \eqref{profileExpansion}.

\subsection{Nonlinear least squares fittings of the Earth--Garand profiles: microwave region}
\label{ssec:garand1}

For temperature sounding in the microwave, a spectral window of less than $1\rm\,cm^{-1}$ is required \citep{Rosenkranz01}.
This is much smaller than the intervals required in the thermal infrared (see below) and contains significantly less spectral lines.
Therefore, we will first test the performance of our approach in the microwave.
To further speed up the simulations we use line parameters of the main molecules from the earlier HITRAN 1986 database \citep{Rothman87}.
For retrieval tests the consistency of the line data used to generate the synthetic observations and the model radiances during the iterative fitting procedure is more important than the
accuracy and completeness of the data base (see also discussion).
To properly account for line wing contributions in the $1.49 \text{--} 2.1 \cm$ window, all lines in this interval extended by $\pm 10 \cm$ are considered:
This gives 835 lines of six molecules (\element{O_2}, \element{H_2O}, \element{O_3}, \element{N_2O}, \element{CO}, \element{CH_4}), whereas the most recent HITRAN 2016 \citep{Gordon17etal}
comprises about 13\,000 lines of the seven ``main'' absorbers (with \element{CO_2} additionally).

Figure \ref{garand_mWave_snr100} illustrates the retrieval performance for the first 18 Garand atmospheres (the AFGL atmospheres and the second group comprising atmospheres with increasing mean temperature).
For the retrievals shown here, a Gaussian S/N=100, and a Gaussian spectral response function with half-width at half-maximum (HWHM) $\Gamma=0.001\cm$ (corresponding to a resolving power $R = \nu / \Gamma \approx 2000$) is considered.
The uniform altitude grid was confined to 26 levels up to $50\rm\,km$.

A comparison of the true temperature and its nonlinear least squares fit (five base vectors) indicates a good agreement (within one to four Kelvin) for altitudes between six and 40\,km (henceforth simply called UTLS, indicating upper troposphere-lower stratosphere), except for some atmospheres exhibiting some zigzag in the mid stratosphere.
Discrepancies increase to several Kelvin near BoA for some cases, and larger deviations up to almost 7\,K at ToA.
The deviations, especially in the upper stratosphere, are clearly evident in \qufig{garand_mWave_snr100}.

For comparison, the temperature profile reconstructed with the linear least squares fit (cf.\ \qeq{linLS}) is shown, too.
In most cases, the pattern of the differences to the true profile is similar, with the linear fit performing better in general.
The linear fit can be seen as a test of the quality attainable by the reconstruction with a small set of basis functions and can be considered as the optimum that can be achieved by fitting
spectra.

An important issue for nonlinear optimization is the initial guess.
Figure \ref{garand_mWave_snr100} also shows that the two choices discussed in the previous Section \ref{ssec:init} yield comparable results:
Except for a few cases starting with the damped mean profile \eqref{initMean} and the straight line \eqref{initLine} leads to roughly the same temperature fit.

\begin{figure}
 \centering\includegraphics[width=\linewidth]{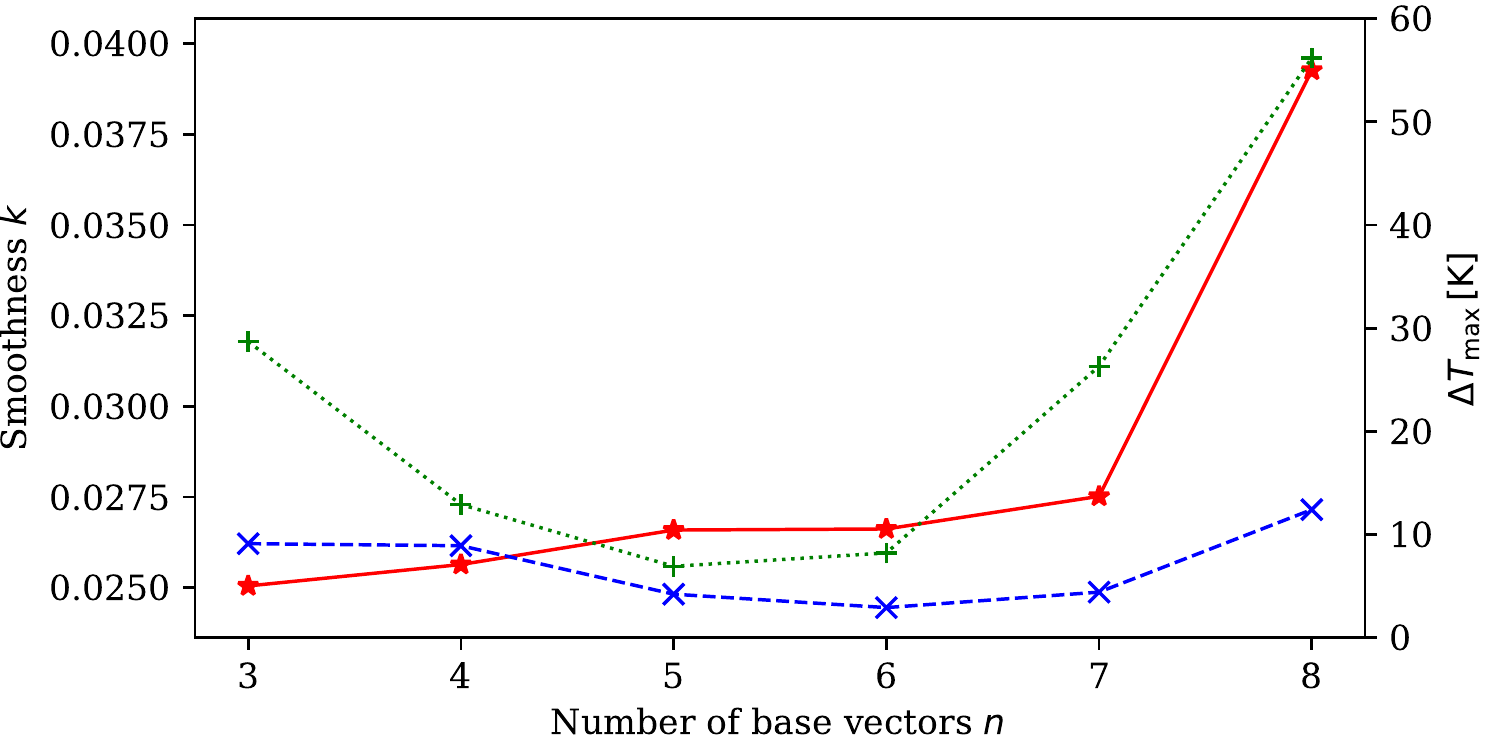}
 \caption{Smoothness $k$ and maximum temperature differences (green dotted: $0 \text{--} 50\rm\,km$, blue dashed: $6 \text{--} 40\rm\,km$) as a function of the number of base vectors.
          Microwave $1.5 \text{--} 2.1\rm\,cm^{-1}$ with SRF $\Gamma=0.001\rm\,cm^{-1}$ and $\rm S/N=5$}
 \label{mWave_nBase} 
\end{figure}

As indicated in Subsection \ref{ssec:linLS}, the number of base vectors determines the quality of the reconstruction:  with too few vectors the fine structure might be lost, whereas with too many
vectors, the zigzags might show up. 
The tests essentially confirm these expectations, with the increasing number of base vectors leading to increasingly stronger oscillations of the error profile [see Appendix].
Moreover, the $n=8$ solutions features strong deviations especially in the upper atmosphere.
The a posteriori parameter choice method \citep{Doicu10} evaluates the sum of consecutive temperature differences to estimate the smoothness,
\begin{equation}
 k = \sqrt{\sum_{l=2}^L (T_l - T_{l-1})^2 \over \sum_{l=1}^L T_l^2} ~,
\end{equation}
where $L$ is the number of altitude grid levels.
Figure \ref{mWave_nBase} indicates that a truncation index 5 or 6 is appropriate.
Moreover, the largest temperature deviation for all 18 atmospheres is minimal with $n=5$ ($6.9\rm\,K$ for the midlatitude winter atmosphere \#3)
and the largest difference in the mid atmosphere ($6 \text{--} 40\rm\,km$) is minimal for $n=6$ ($2.9\rm\,K$ for the subarctic summer atmosphere \#4).
Henceforth five base vectors will be used for all retrievals.

\begin{figure*}
 \centering\includegraphics[width=\textwidth]{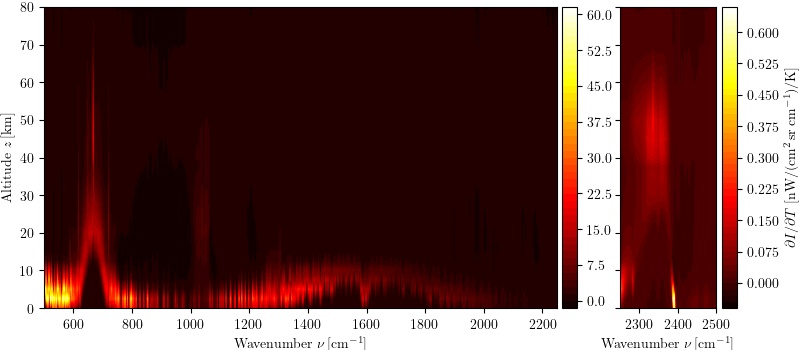}
 \caption{Temperature Jacobian for an Earth midlatitude summer atmosphere (Gauss response with HWHM $\Gamma=1\cm$).}
 \label{tirJacobian} 
\end{figure*}

The retrievals so far have been performed for synthetic observations with Gaussian noise added corresponding to an S/N 100.
However, this might be too optimistic for real exoplanet spectra.
For doubled and quadrupled noise, the deviations stay within a few Kelvin in the UTLS, but for $\rm S/N=10$ larger deviations in the upper stratosphere become apparent (worst case $\Delta T \approx 19\rm\,K$ at ToA of atmosphere \#10).
Nevertheless, the UTLS error never grows larger than $10\rm \,K$.

Besides noise, the resolution of the instrument is expected to have an impact on the quality of the retrievals.
Doubling the half width to $\Gamma=0.002\cm$ does not significantly deteriorate the quality of the fits in the UTLS;
however, a half width $\Gamma=0.004\cm$ corresponding to a resolving power $R=500$ significantly worsens the fits, with the largest errors again in the upper atmosphere
(worst case $\Delta T \approx 29\rm\,K$ at ToA of the coldest atmosphere \# 7).

\begin{figure*}
 \centering
\includegraphics[width=\textwidth]{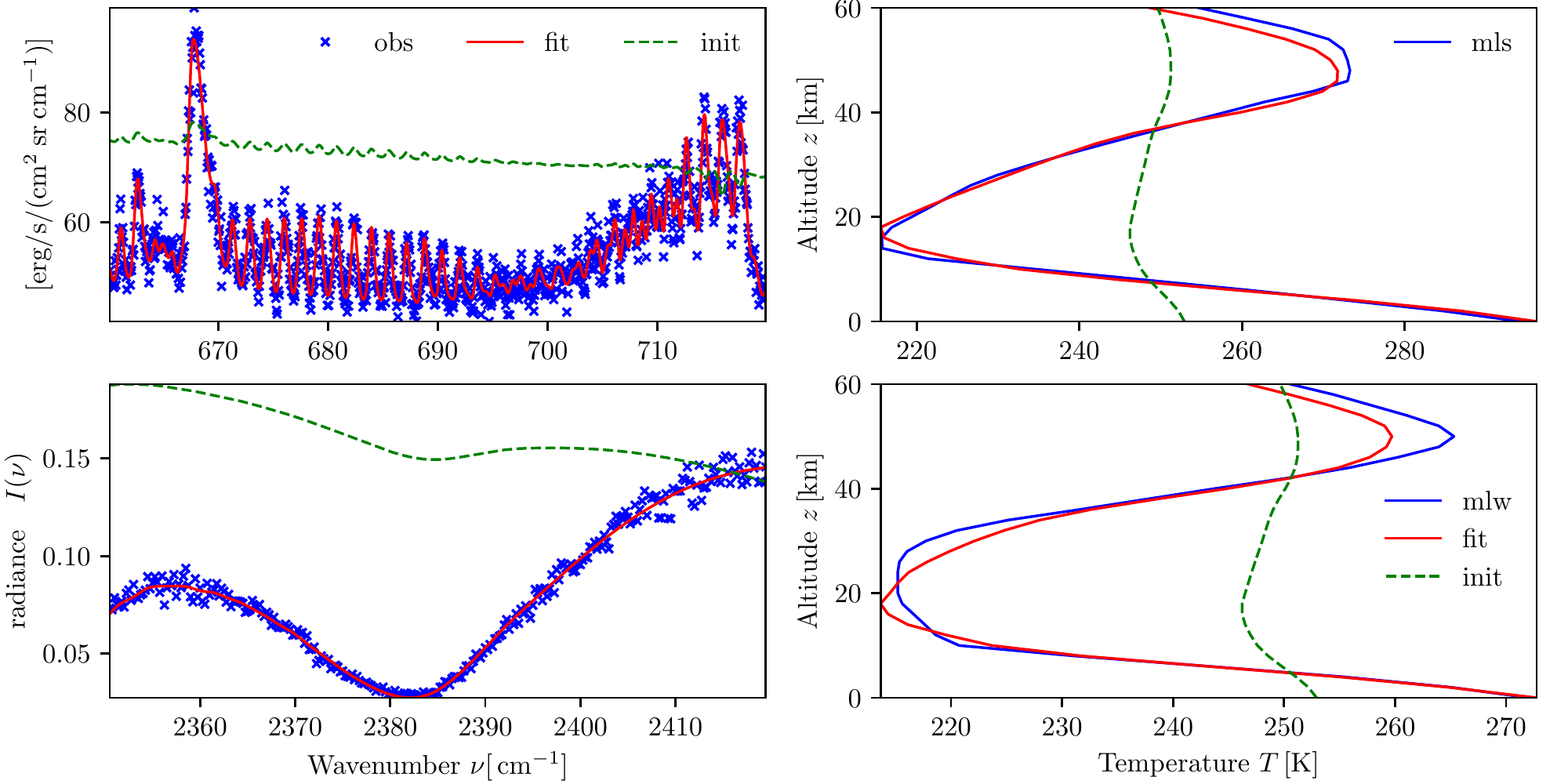} 
 \caption{Comparison of observed, final fit, and initial guess spectra (left) and true, fitted, and initial guess temperature profiles (right).
          The initial guess temperature profiles are identical in both cases.
          Top: TIR $15\mue$ for midlatitude summer (mls), Gaussian response with $\Gamma=0.25\cm$. Bottom: TIR $4\mue$ for midlatitude winter (mlw), Gauss $\Gamma=1.0\cm$; both with S/N=25.}
 \label{tirSpectra}
\end{figure*}

\subsection{Nonlinear least squares fit of the Earth--Garand profiles: TIR}
\label{ssec:garandTIR}

For the remote sensing of atmospheric state parameters, the selection of suitable spectral regions is mandatory and numerous studies on ``microwindow'' or ``channel'' selection methods have been presented
\citep[e.g.,][]{Clarmann98,Rabier02}.
In addition to an analysis of the weighting functions (cf.\ \qufig{wgtFct_tir2} and \qufig{wgtFct_tir3}), inspection of the Jacobian $\partial I(\nu) / \partial T(z)$ is particularly useful.
Figure \ref{tirJacobian} indicates high sensitivity to temperature in the $\nu_2$ and $\nu_3$ bands of \element{CO_2} up to the mesosphere.
The AIRS (Atmospheric Infrared Sounder) grating instrument (with a resolution of $R=\nu/\Delta\nu =1200$) aboard the NASA polar-orbiting satellite AQUA essentially exploits the $664 \text{--} 713\cm$ and $2197 \text{--} 2395\cm$ region (in addition further channels scattered through the TIR are used),
whereas the IASI (Infrared Atmospheric Sounding Interferometer) Fourier transform spectrometer (with spectral resolution 0.25\cm) aboard the ESA/EUMETSAT MetOp satellites considers the $650 \text{--} 770\cm$, $1210 \text{--} 1650\cm$, $2150 \text{--} 2250\cm$, and $2350 \text{--} 2420\cm$ windows \citep[see also][]{Menzel18}.

\begin{figure*}
 \centering
\includegraphics[viewport=0 0 864 550, clip=true, width=\textwidth]{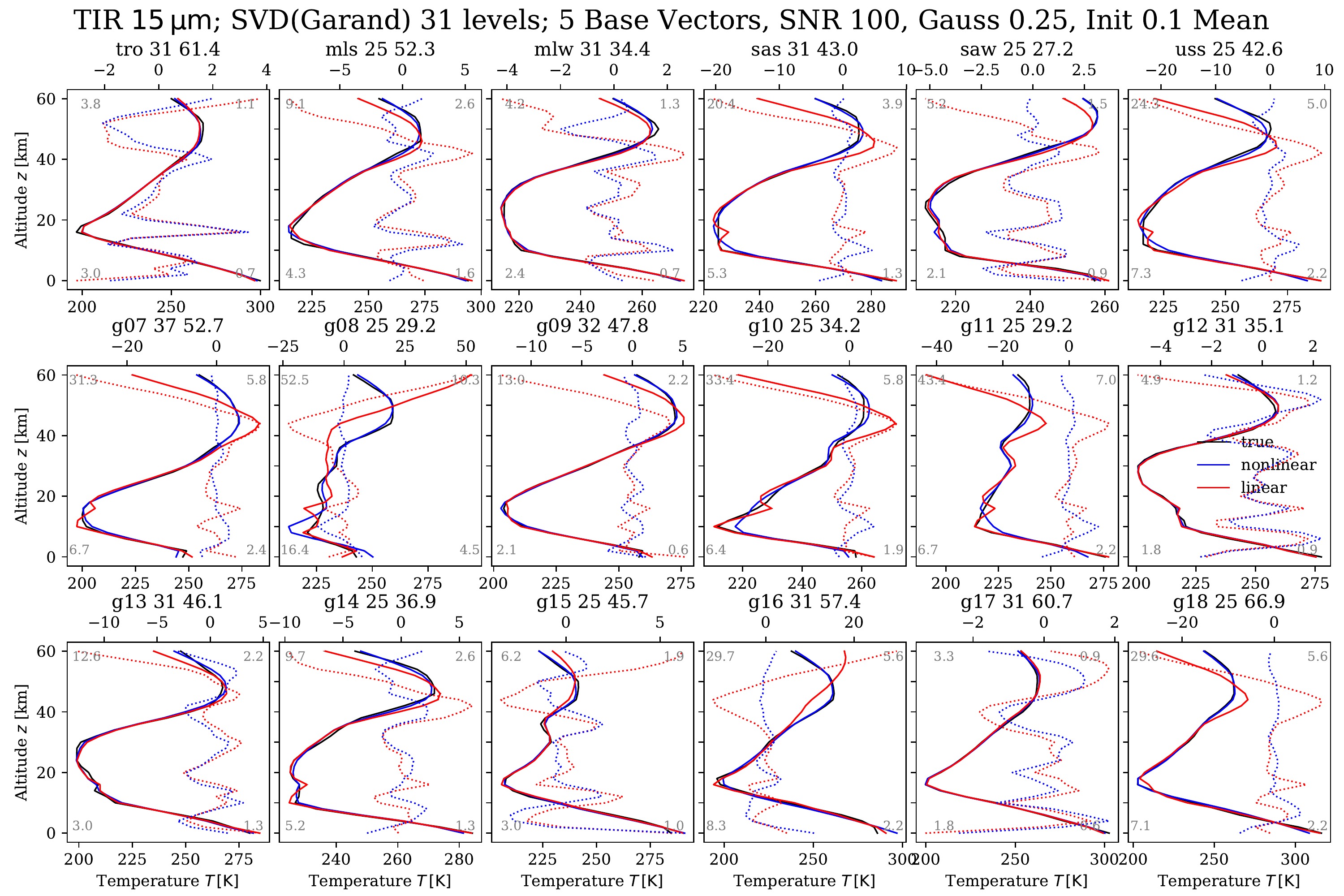} 

\includegraphics[viewport=0 0 864 550, clip=true, width=\textwidth]{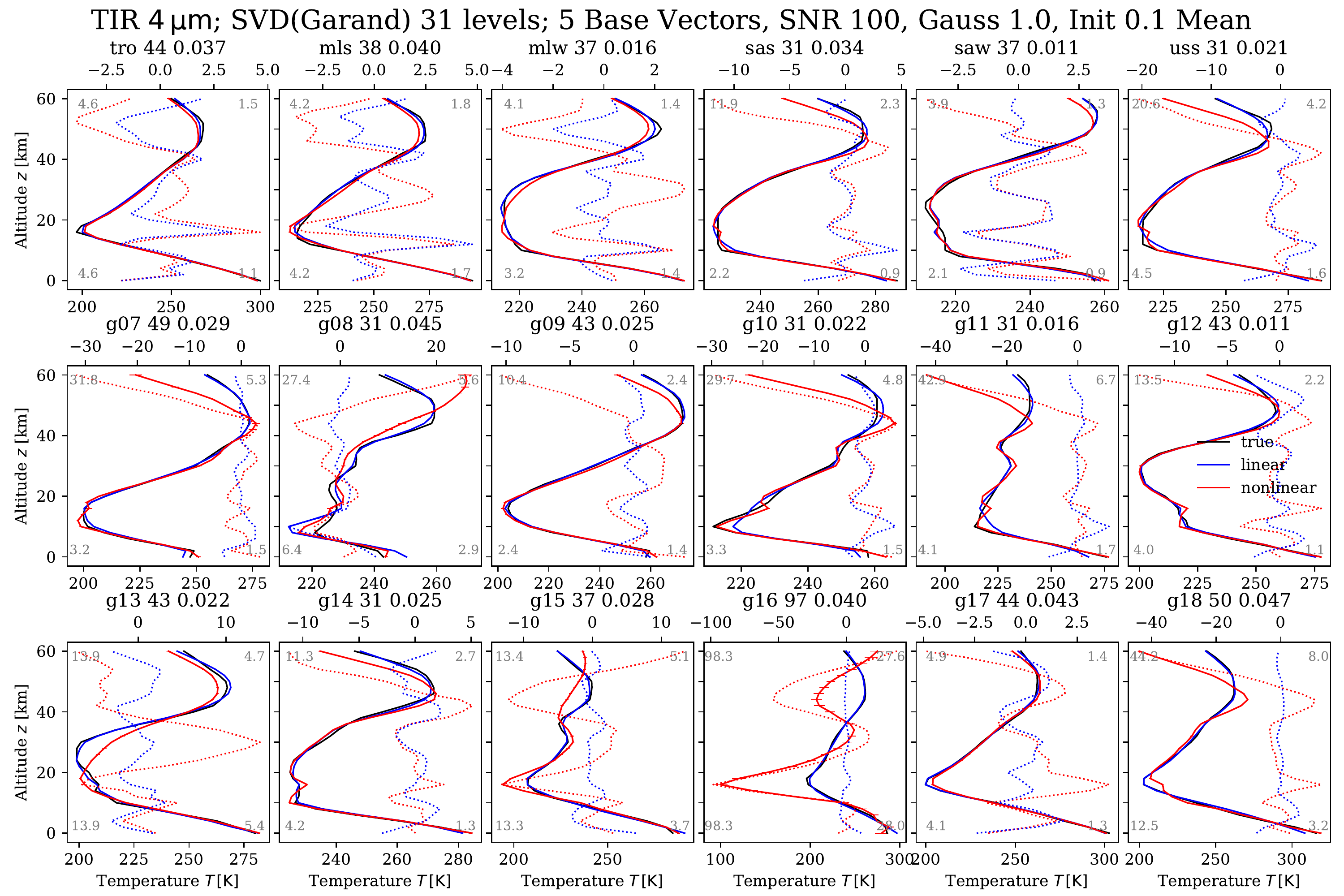}
 \caption{TIR temperature retrievals for the first 18 Garand atmospheres with SVD of the Garand ``temperature matrix'' (S/N=100 and Gaussian SRF).
          Five base vectors, initial guess damped mean temperature \eqref{initMean}.
          Top: LW $660 \text{--} 730\cm$ ($15\mue$) with HWHM $\Gamma=0.25\cm$ (resolution $R \approx 2800$);
          bottom: SW $2350 \text{--} 2420\cm$ ($4.3\mue$) with $\Gamma=1.0\cm$ ($R \approx 2400$).
          Line colors and styles and title as in \qufig{garand_mWave_snr100}.
          }
 \label{garand_tir_svdE_snr100}
\end{figure*}

\begin{figure*}
 \centering
\includegraphics[viewport=0 0 864 550, clip=true, width=\textwidth]{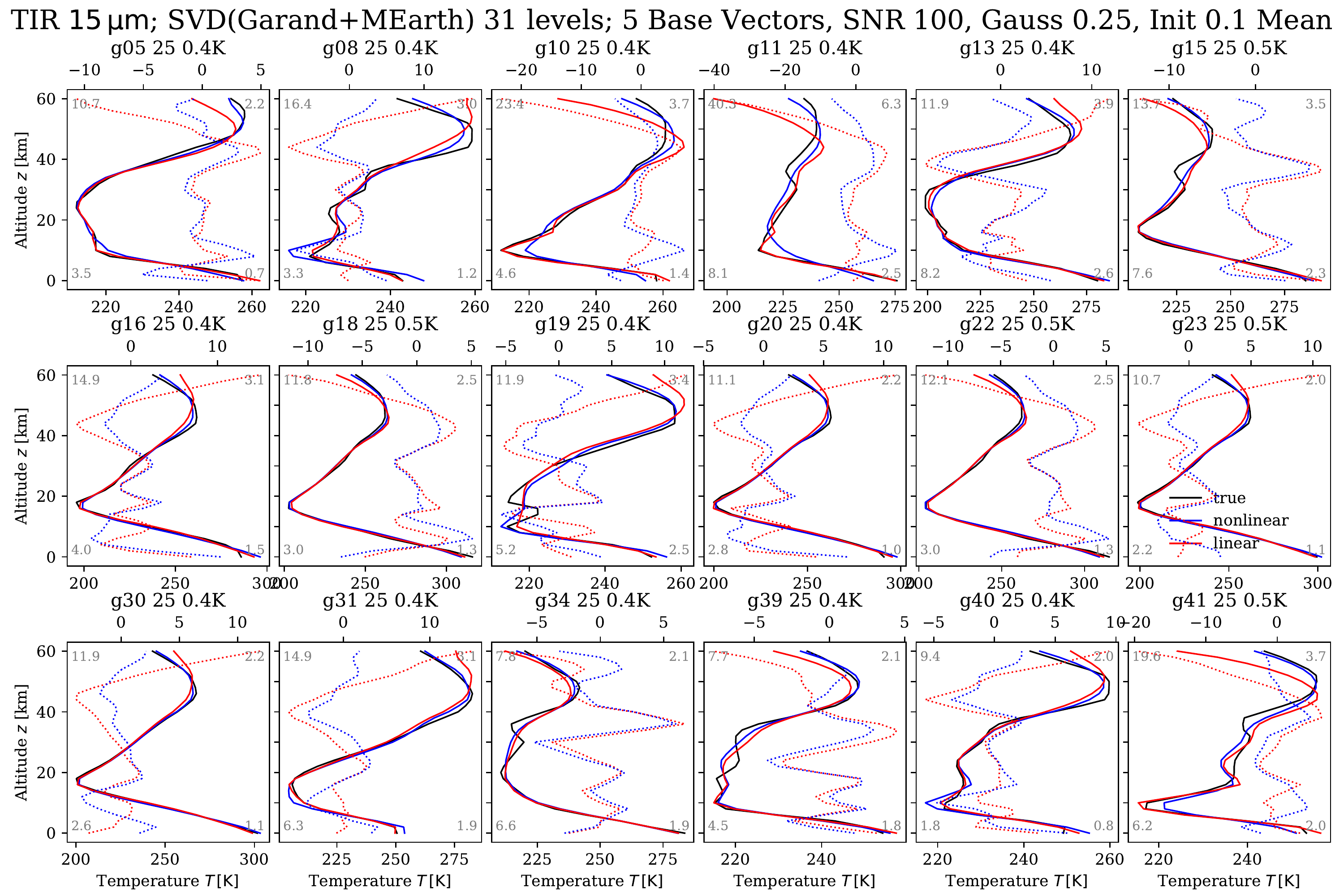}
 \includegraphics[viewport=0 0 864 550, clip=true, width=\textwidth]{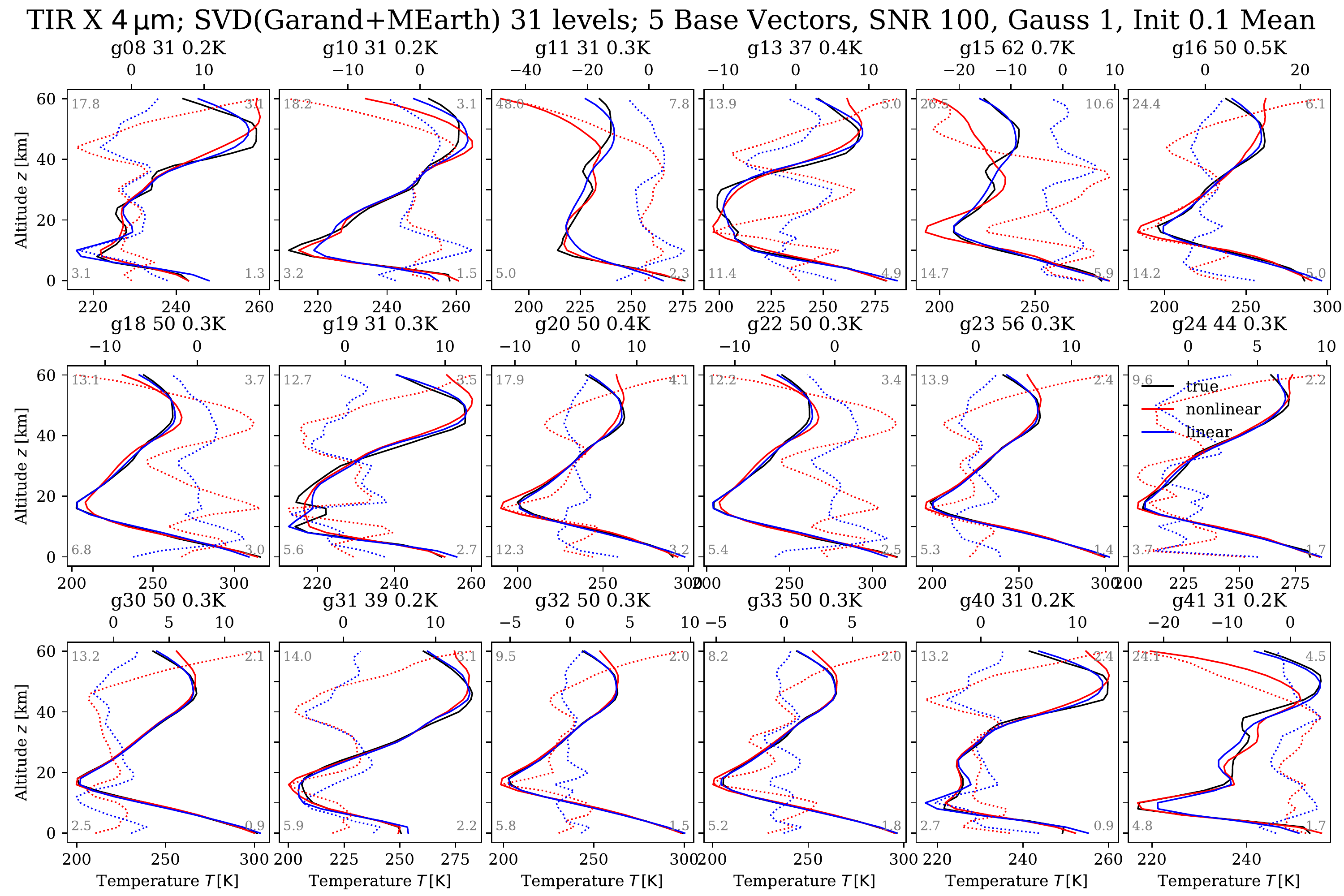}
 \caption{TIR temperature retrievals for the Garand atmospheres with SVD of the combined Garand+MEarths temperature matrix (S/N=100 and Gaussian SRF as above).
          Top: LW $660 \text{--} 730\cm$, poor fits with $|\Delta T| > 7.5\rm\,K$;
          bottom: $2200 \text{--} 2400\cm$, only fits with $\Delta T \ge 8\rm\,K$.
          The titles give the atmosphere, number of iterations, and the mean brightness temperature difference.
          For further details, see \qufig{garand_tir_svdE_snr100}.}
 \label{garand_tir_svdEM_snr100}
\end{figure*}

Figures \ref{wgtFct_tir2} and \ref{wgtFct_tir3} clearly show that resolution impacts the sensitivity to the upper atmosphere.
For high resolution (Gaussian response with $\Gamma=0.01\cm$), the $\nu_2$ band around $15\mue$ is sensitive up to $50\rm\,km$, and the $\nu_3$ band at $4.3\mue$ even higher.
With a Gauss of width $1\cm$ the sensitivity for the upper stratosphere is damped especially in the longwave regime.
Both the Mid-Infrared Instrument (MIRI) medium-resolution spectrometer (MRS) and the Near Infrared Spectrograph (NIRSpec) instrument on JWST will offer a resolution of $R \approx 2500$, hence, we consider a Gaussian response function with HWHM $\Gamma=0.25\cm$ and $\Gamma=1.0$ for the TIR $\nu_2$ and $\nu_3$ bands centered roughly around $700\cm$ and $2400\cm$, respectively.

Longwave and shortwave TIR spectra and the corresponding fitted temperature profiles are shown in \qufig{tirSpectra}.
Despite the poor initial guess temperature \eqref{initMean} and the almost flat corresponding radiance spectrum the least squares fit performs quite well, with model spectra essentially consistent with the noisy observations and fit temperatures close to the truth.

\subsubsection{Moderate noise}

Figure \ref{garand_tir_svdE_snr100} shows the temperature retrievals for the TIR-SW and TIR-LW bands with an admittedly very optimistic noise level of $\text{S/N}=100$.
Similar to the microwave retrievals, deviations show up especially at BoA and are usually more pronounced at ToA.
In the UTLS the mean deviation is never larger than five Kelvin and the maximum deviation is rarely larger than ten Kelvin for the $15\mue$ retrievals; for almost one half of the fits (7 of 18) the maximum difference
is larger than $5 \rm\, K$.

The performance of the $4.3\mue$ retrievals is somewhat worse: in particular, the retrieval fails for the Garand \# 16 atmosphere which has the highest and coldest tropopause
(the number of iterations is also exceptionally high).

The choice of the set of profiles considered in the ``temperature matrix'' used for the SVD (cf.\ Section \ref{ssec:base}) does not have a significant impact on the fit.
The residual norms $\| \vec y \|$ are roughly equivalent for both spectral regions (with the exception of atmosphere \# 8 in the SW-TIR, where the norm differs by more than a factor 2).
Furthermore, the number of iterations is often identical.
Using the base vectors resulting from the decomposition of the combined 42 Garand and 11 M-Earth profiles is likely able to capture more atmospheric scenarios and will be used in the following.
Figure \ref{garand_tir_svdEM_snr100} shows the worst 18 fits for all 42 atmospheres: we note that 14 of the atmospheres (i.e.,\ one third) appear to be ``problematic'' for both TIR regions.

Similar to the previous SW TIR runs (\qufig{garand_tir_svdE_snr100}, bottom), the fit of the Garand \#16 atmosphere also failed with the new set of base vectors.
Moreover, the fitted Garand \#31 temperature (the warmest stratopause at a very low height) is also significantly different from the true temperature (deviation almost $100 \rm\, K$).
In both cases the linear least squares fit \eqref{profileExpansion} and the fit of the LW-TIR spectra perform quite well, hence the number of base vectors cannot explain the problems with these two atmospheres.
Inspection of the SW TIR weighting functions (\qufig{wgtFct_tir3}) suggests that the $2350 \text{--} 2420\cm$ spectral window may be too small.
In fact, this spectral interval has little sensitivity to the lower troposphere, and retrievals with an extended interval $2200 \text{--} 2420\cm$ show better results (\qufig{garand_tir_svdEM_snr100}).
However, this significantly increases the computational run time.  

\begin{figure*}
 \centering
\includegraphics[viewport=0 0 864 550, clip=true, width=\textwidth]{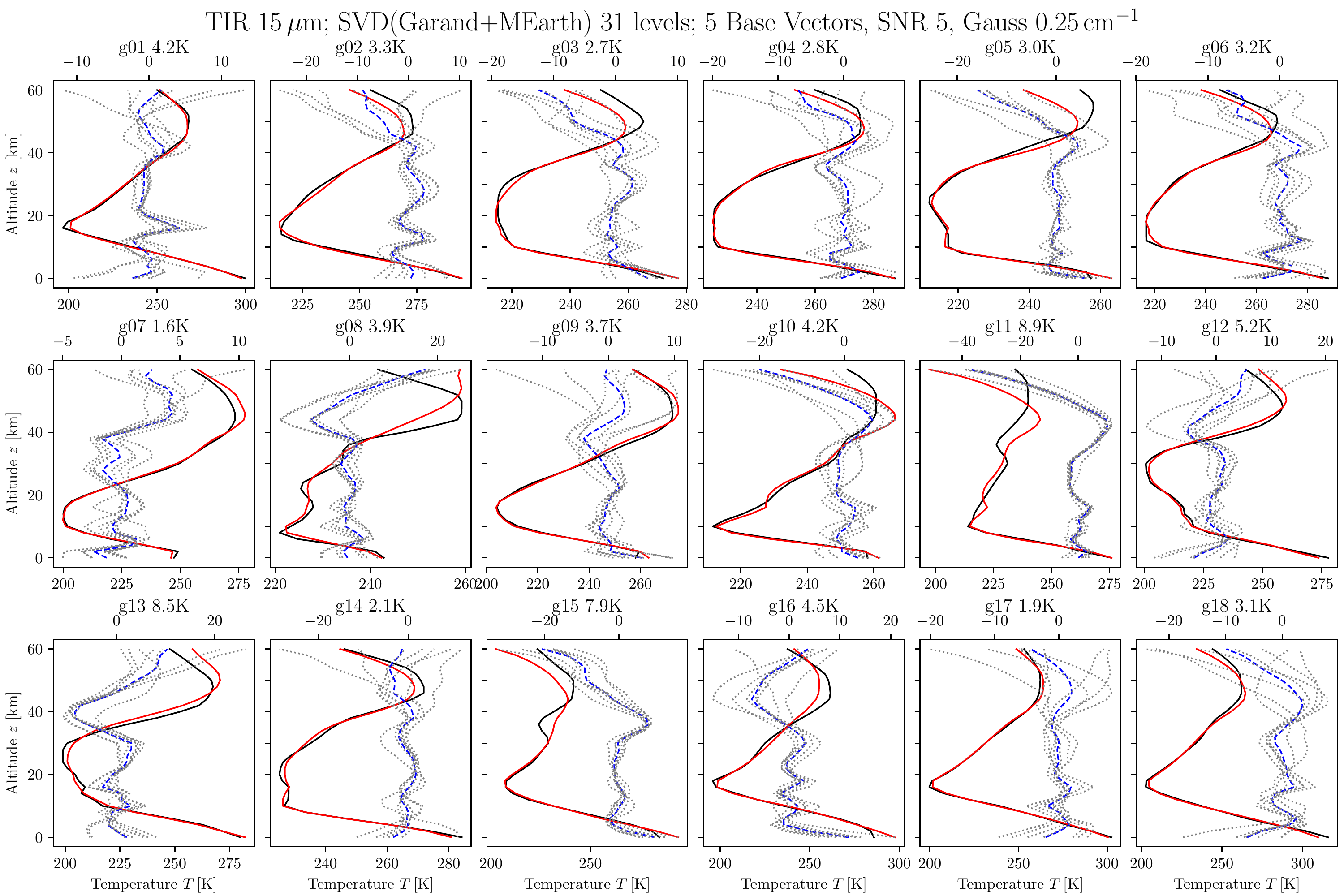}
 \caption{LW TIR $660 \text{--} 720\cm$ ($15\mue$) temperature fit for 18 Garand atmospheres with SVD of the combined Garand+MEarths temperature matrix: Gauss SRF with S/N=5 and HWHM $\Gamma=0.25\cm$ ($R \approx 2800$).
          Black: true; red: mean fit; blue: error ($\Delta T$) of mean fit; gray: errors of individual fits with different initial guesses.
          The number in the title is the largest error in the UTLS.}
 \label{tir2_garand_5base_snr005} 
\end{figure*}

\subsubsection{Strong noise}

The S/N=100 spectra analyzed so far are extremely optimistic for exoplanet observations but they have served to demonstrate the feasibility of the retrieval approach.
As a more realistic scenario we consider synthetic measurement spectra with an S/N=5; this is probably out of reach with current or near future instrumentation for a single measurement, but it might be
obtainable by co-adding several observations.
As expected, the differences to the true profiles for a single fit are larger than for the S/N=100 case (\qufig{garand_tir_svdEM_snr100}).
The estimated temperature profile changes with different noise vector realizations, but the average of several independent fits is closer to the truth.
However, in reality, there will be only a single measured noisy spectrum.

Different initial guess profiles will also yield different estimates of the temperature
and, accordingly, the fit has been repeated for a series of initial guesses, the damped mean profile \eqref{initMean}, the straight line \eqref{initLine}, and linear combinations thereof.
In Fig.\ \ref{tir2_garand_5base_snr005} the results of least squares fits of $15\mue$ observations is shown, clearly demonstrating that the mean temperature obtained by averaging gives a good fit:
for some fits, the differences can be as large as 20\,K in the upper stratosphere.
However, similar to the S/N=100 case, the fits are reasonably good in the UTLS region, with the maximum error of almost 9\,K (atmosphere \# 11).
In the $2 \text{--} 30\rm\,km$ region the maximum error is never larger than 4.2\,K (atmosphere \# 1).

Least squares fits of the SW TIR observations ($4\mue$) are in general worse, for one third of the 18 cases the deviation of the averaged temperature fit is larger than 10\,K in the UTLS.
Nevertheless, in all cases some of the individual fits are acceptable, and the average of several fits can be used as for the LW case.


\subsection{Least squares fits of Earths orbiting M-dwarfs}
\label{ssec:MEarths}

The previous subsections have clearly demonstrated that a least squares fit of function expansion coefficients can be used successfully to infer Earth's atmospheric temperature profiles.
Our main objective is the characterization of exoplanetary atmospheres, and in this subsection we continue with an analysis of M-dwarf planet atmospheres.

\begin{figure*}
 \centering\includegraphics[width=\textwidth]{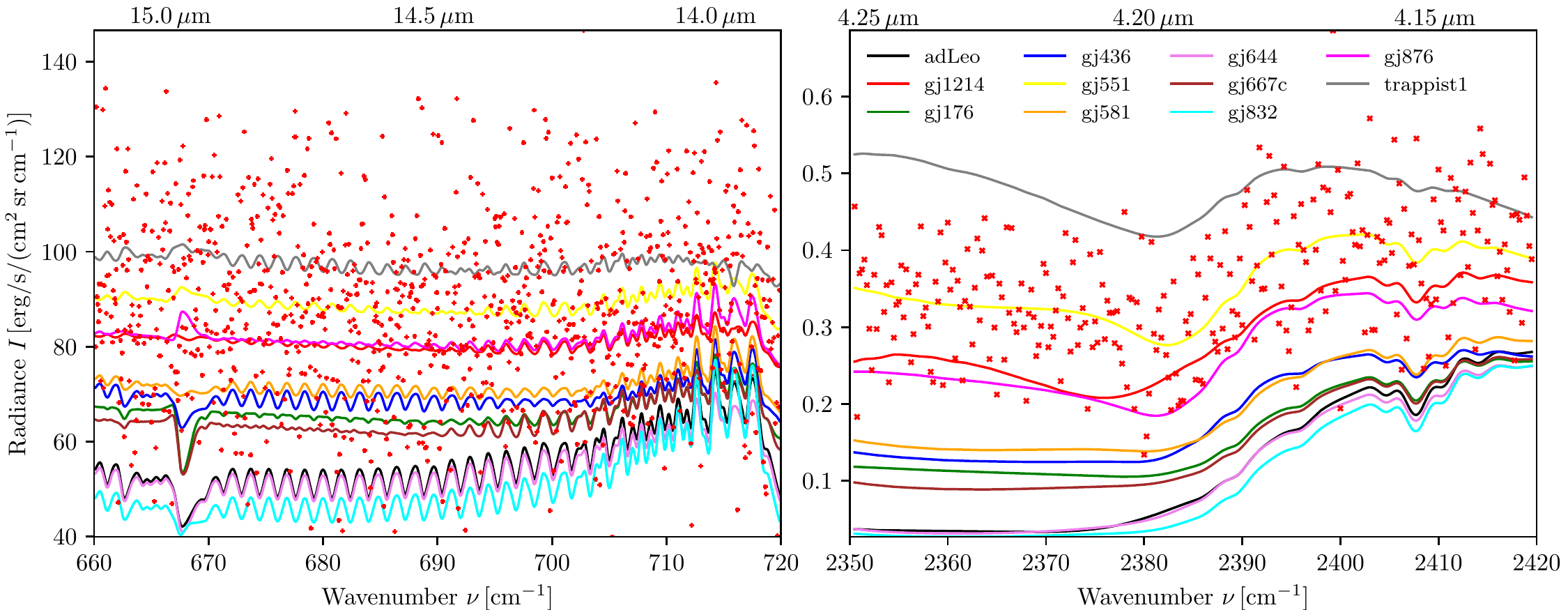}
 \caption{Longwave (left) and shortwave M-Earth TIR spectra. Gaussian spectral response with $R \approx 2800$ (LW) and $R \approx 2400$ (SW).
  Solid lines show the fitted model spectra. Furthermore, the noise (S/N=5) contaminated ``observation'' of GJ 551 (Proxima Centauri) is shown (red points).} 
 \label{MEarth-TIR-spectra} 
\end{figure*}

The temperature matrix comprised of the eleven M-Earth profiles is too small to generate a (left) singular matrix useful for a representative set of base vectors.
On the other hand, the temperature matrix also including the Garand profiles (i.e.,\ 42+11 profiles) proved to be useful for analysis of the synthetic Earth spectra
(cf.\ \qufig{garand_tir_svdEM_snr100} and \ref{tir2_garand_5base_snr005}) and the base vectors resulting from its SVD are, hence, to be used here as well.

\begin{figure*}
 \centering
\includegraphics[viewport=0 0 864 409, clip=true, width=\textwidth]{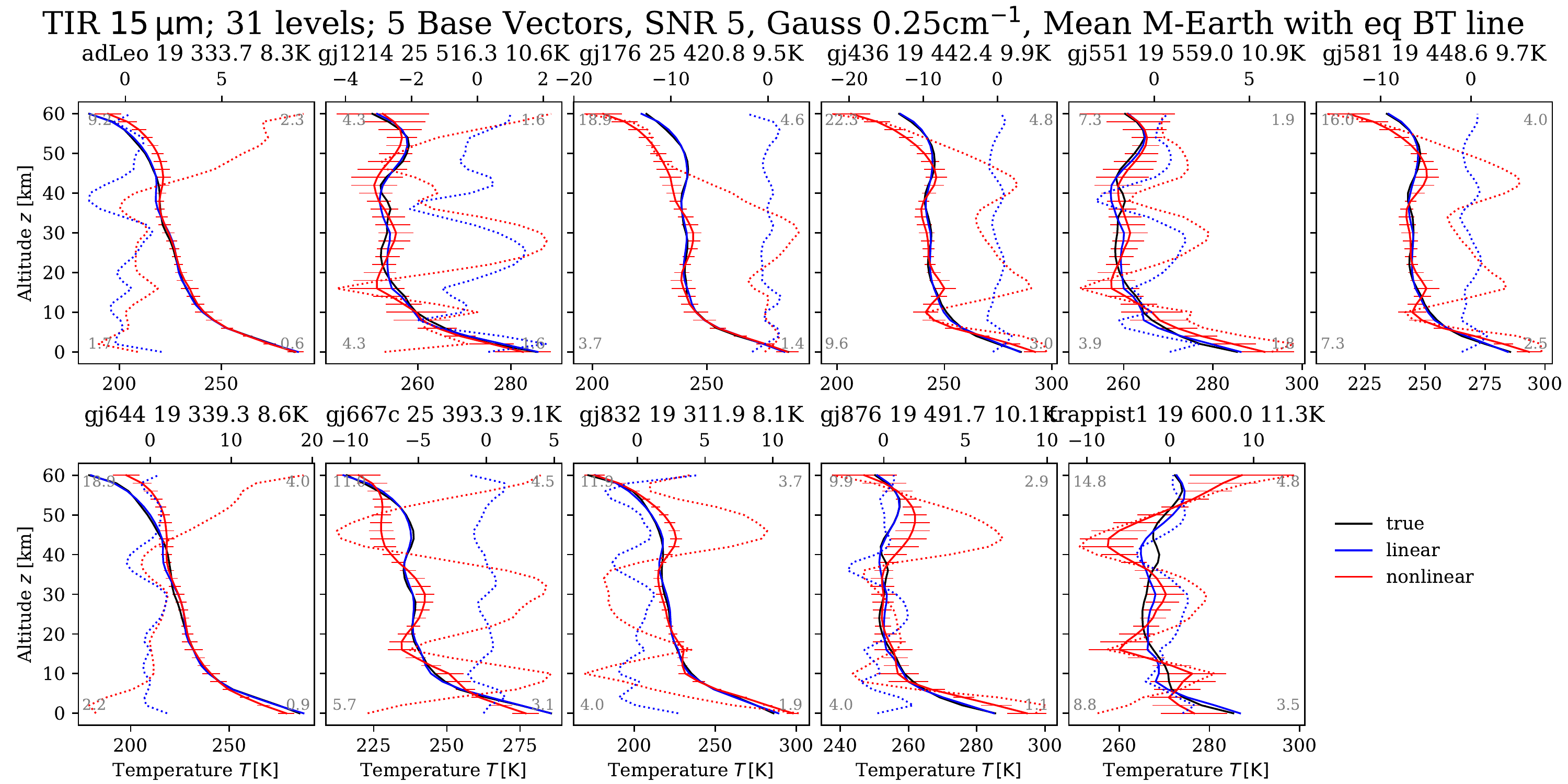} 

 \includegraphics[viewport=0 0 864 409, clip=true, width=\textwidth]{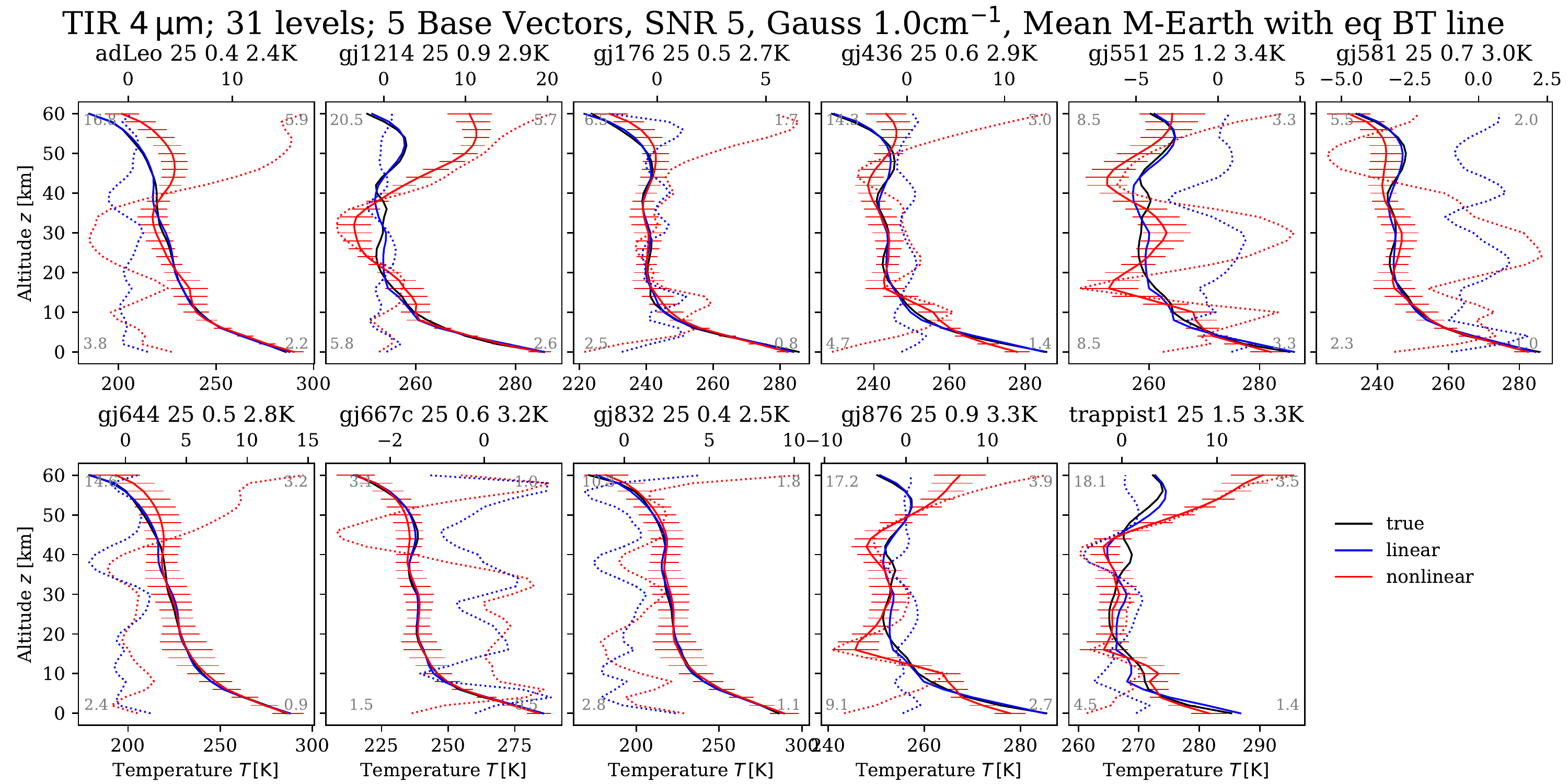}
 \caption{Nonlinear least squares fit of the Earths around M-dwarfs using the mean M-Earth atmosphere in the forward model.
          S/N=5.
          Top: LW TIR band with $R\approx 2800$,
          Bottom: SW TIR band with $R\approx 2400$.
          Numbers in the corners are deviations as in \qufig{garand_tir_svdE_snr100} and \ref{garand_tir_svdEM_snr100}.
          The title gives the number of iterations, the residuum norm, and the mean brightness temperature difference. Line colors and styles are as in \qufig{garand_mWave_snr100}.
          }
 \label{MEarth-TIR} 
\end{figure*}

Temperatures retrieved from noisy (S/N=5) synthetic M-Earth thermal IR spectra (see \qufig{MEarth-TIR-spectra}) are shown in \qufig{MEarth-TIR}. Here mean M-Earth pressure and concentration profiles have been used in the forward model.
The initial guess is defined by \eqref{initLine} with the ToA and BoA temperature given by the minimum and maximum observed equivalent brightness temperature.
In contrast to the Earth retrievals, both the shortwave and longwave TIR appear to work equally well:
the largest temperature error ($\Delta T$) is 20\,K in the upper atmosphere, but less than 10\,K in the UTLS region.%
\footnote{We continue to use this term for the $6 \text{--} 40 \rm\,km$ region although some of the planets do not have a stratosphere similar to that of the Earth.}
However, error bars are significantly larger at this point, indicating uncertainties up to some $10\rm\,K$.
Moreover, significant deviations in the upper atmosphere can be noticed for about half of the planets.

The analysis of exoplanet observations is generally more challenging than for Earth (although accuracy requirements for Earth remote sensing are usually significantly higher and time
constraints for operational processing of thousands or millions of observations are not uncommon).
For exoplanets, the composition of the atmosphere is largely unknown and the concentration of \element{CO_2} or \element{O_2} is not necessarily constant with altitude as for Earth.
In particular, for the M-Earths, the \element{CO_2} VMR is constant at 355\,ppm in the lower atmosphere up to about 15\,km and then increases (for some planets to more than 500\,ppm at ToA).

\begin{figure}
 \centering\includegraphics[width=\linewidth]{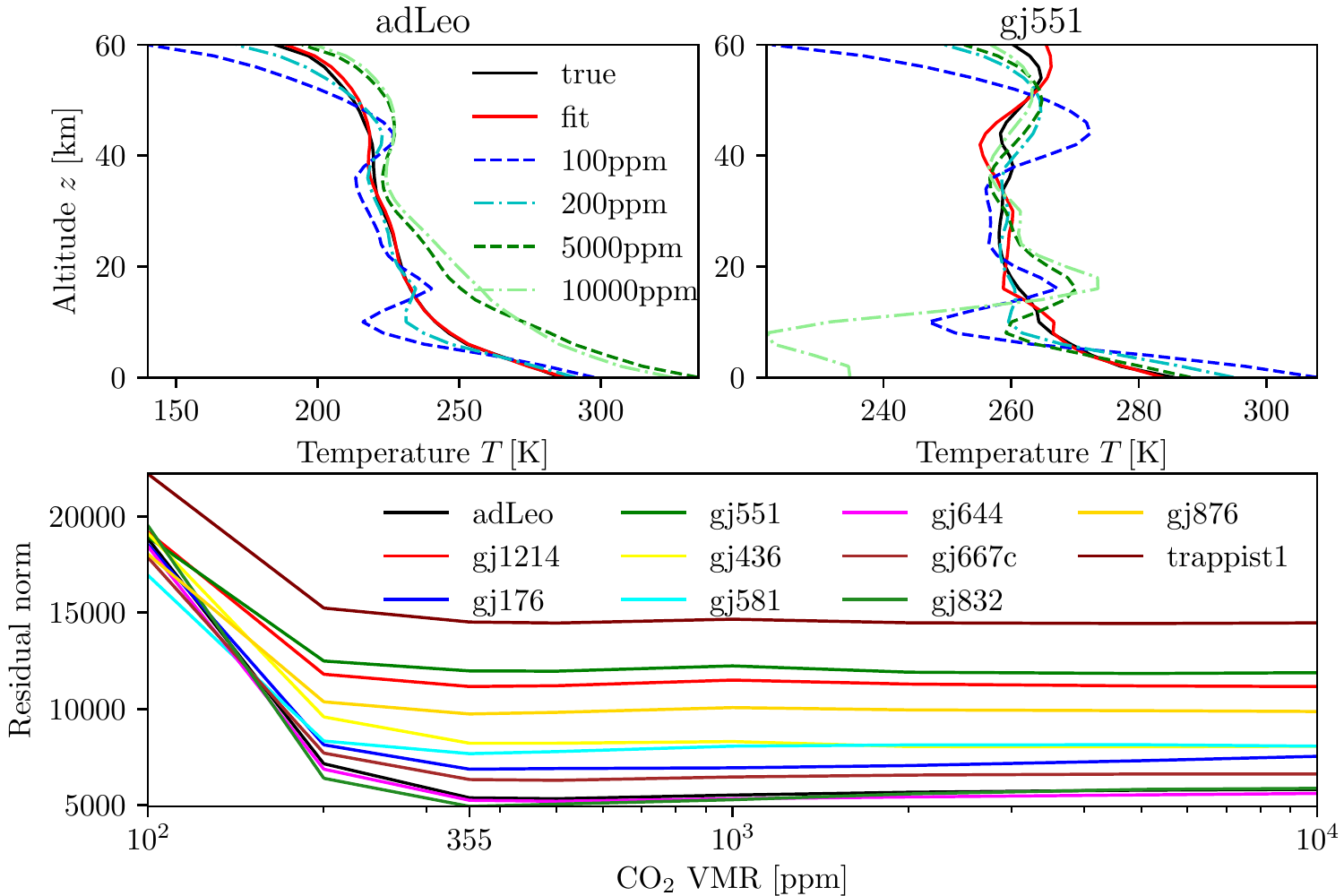}
 \caption{Nonlinear least squares fit of the M-Earths using the mean M-Earth atmosphere with different \element{CO_2} mixing ratios (LW TIR, S/N=25).
          Top: solid red line shows the temperature retrieved with the correct, altitude-dependent \element{CO_2} profile in the forward model.
          Bottom: residual norm as a function of the \element{CO_2} VMR (the value at 355\,ppm corresponds to the fit with the correct \element{CO_2}).} 
 \label{MEarth-TIR-CO2} 
\end{figure}

Retrieval tests with $355\rm\,ppm$ \element{CO_2} throughout the entire atmosphere or with a somewhat increased (390\,ppm) or decreased (320\,ppm) isoprofile performed slighty worse,
but in the UTLS region, the temperature profile is still reproduced reasonably well.
Further reducing \element{CO_2} in the forward model data significantly deteriorates the retrieved temperature,
particularly for 200\,ppm, as the profile starts to oscillate, and the zigzag behaviour is even worse for 100\,ppm; see \qufig{MEarth-TIR-CO2} (top).
The retrievals appear to be more tolerant with high partial pressures of \element{CO_2}; temperatures in the lower and mid atmosphere are often overestimated and somewhat smoother;
in some cases the temperature is underestimated in the troposphere and oscillating above.
Moreover, the error bars can become conspicuously large.

However, the true temperature is unknown for real observations and the only hint on the fit quality is the similarity or mismatch of the observed and modeled spectrum.
Indeed, the norm of the residual $\vec{I}_\text{obs} - \vec{I}_\text{mod}$ is significantly larger for 100\,ppm \element{CO_2} and somewhat larger for 1000\,ppm \element{CO_2} compared to the fit with the correct VMR,
cf.\ \qufig{MEarth-TIR-CO2} (bottom).

As discussed in Section \ref{ssec:data}, an actual M-Earth atmosphere can differ substantially from the mean M-Earth atmosphere
and the use of this mean in the forward model might be problematic.
For an assessment of the impact of the model atmosphere used in the nonlinear least squares solver the synthetic S/N=5 spectra of the eleven M-Earths have been fitted assuming each of the M-Earth
atmosphere models.
In the case of real observations the true model atmosphere will be unknown, and a series of fits using different models might be a reasonable approach to tackle the problem.
The ``final'' temperature profile can then be selected as the result of the ``best'' fit (e.g.,\ characterized by the smallest residual norm) or the mean of the individual temperature fits
(optionally weighted by the ``quality'' of the fit, i.e.,\ residuum norm).
The initial guess profile is defined by a straight line \eqref{initLine} with $T_\text{BoA}$ and $T_\text{ToA}$ given  by the minimum and maximum observed brightness temperature. 

\begin{figure*}
 \centering\includegraphics[viewport=0 0 864 409, clip=true, width=\textwidth]{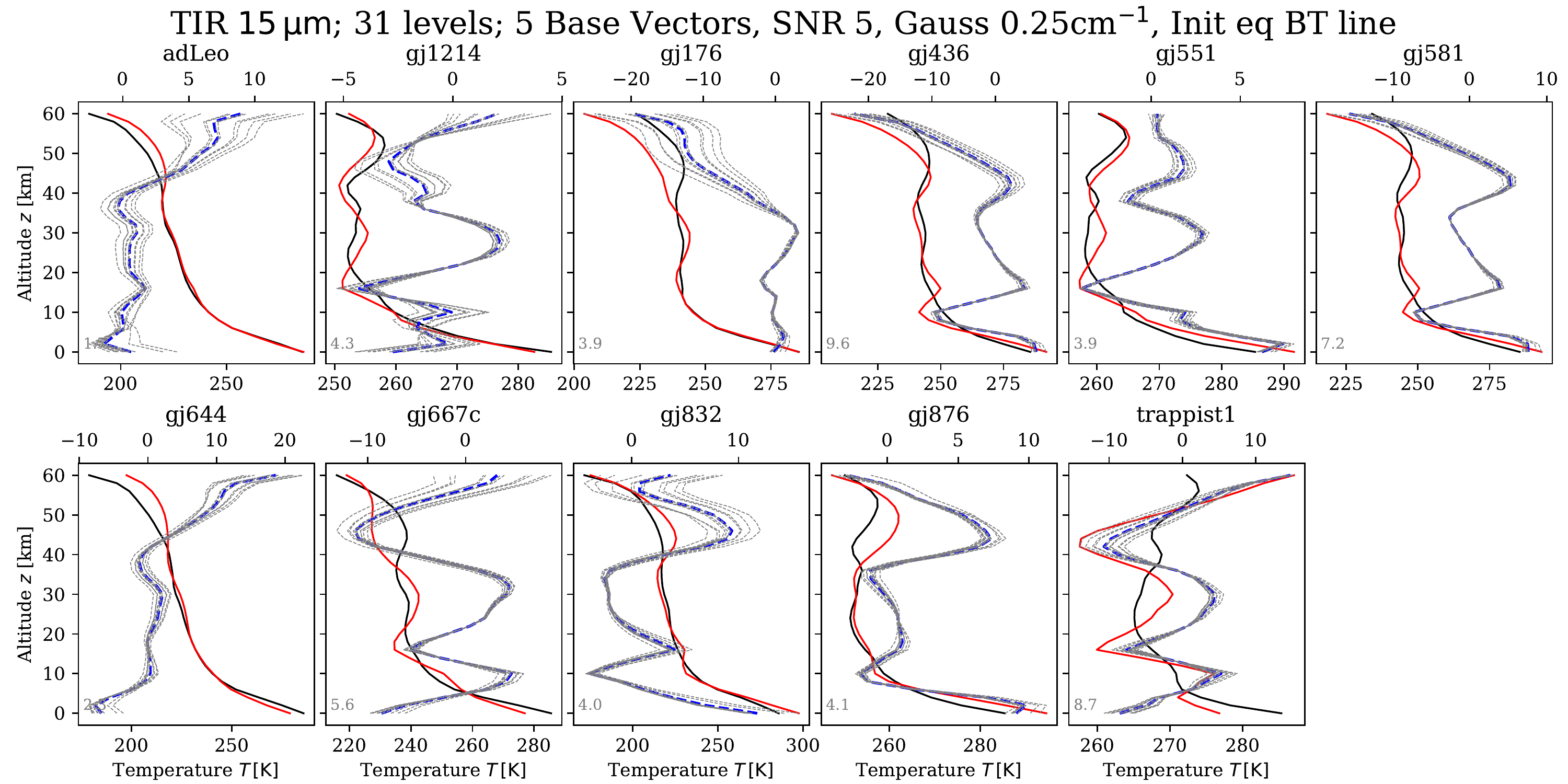} 
 \centering\includegraphics[viewport=0 0 864 409, clip=true, width=\textwidth]{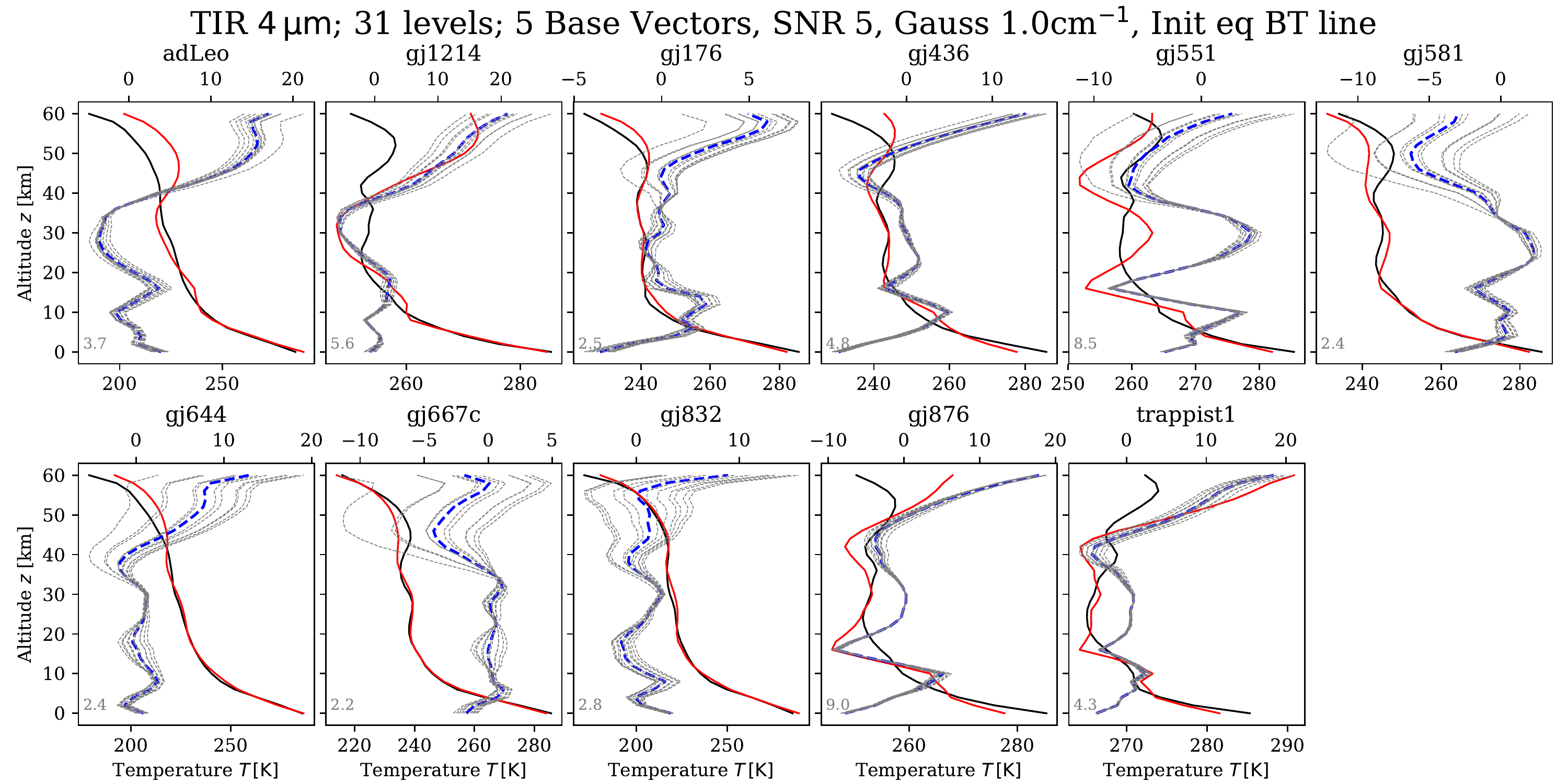}
 \caption{Nonlinear least squares fit of the Earths around M-dwarfs: Impact of model atmospheres.
          Top: LW TIR band with $R\approx 2800$. Bottom: SW TIR band with $R\approx 2400$.
          Black: true; red: mean fit; blue: error ($\Delta T$) of mean fit; gray: errors of individual fits.
          }
 \label{MEarth-TIR-models} 
\end{figure*}

Figure \ref{MEarth-TIR-models} summarizes the results of these $11 \times 11$ fits in the short- and longwave TIR.
Here the final temperature profile is obtained by the mean of the eleven profiles.
Similar to the fits shown above, the temperature often shows large deviations in the upper atmosphere, but in the UTLS region the deviations from the true profile are in the range of a few Kelvin,
with a maximum deviation of up to 10\,K.

\subsection{Least squares fits of Earths orbiting G and K stars}
\label{ssec:FGK}

Figure \ref{atmosTemp} indicates that the atmospheric data of some of the G-star planets are limited in their altitude range.
In order to avoid the extrapolation of profiles, only G-star planets with a ToA altitude $\ge 60\rm\,km$ were considered (i.e., only 44 of the 56 data sets).
Similarly as with the Garand-Earth data and the M-Earth data, the atmospheres have been regridded to a uniform altitude grid of 31 levels with $\delta z = 2\rm\,km$ steps
and the temperature vectors are stacked as columns in a ``G-temperature matrix''.

The base vectors resulting from the SVD of the combined Garand-Earth and M-Earth temperature matrix have been useful for analysis of both the Earth and M-Earth spectra and, therefore, they have also been
used for the analysis of the synthetic G-star planet observations.
However, for many of these fits the deviations of the fitted to the true temperature are large:
the LW-TIR fits are somewhat better; here only five fits have a maximum error in the UTLS larger than $10\rm\,K$ Kelvin (maximum $13.7\rm\,K$),
whereas in the SW-TIR seven fits have maximum UTLS errors from $10 \text{--} 20\rm\,K$, and the GS103RHMWA0.22 case fails completely.
This suggests that these base vectors are not flexible enough to cover typical G-star planetary temperatures.

\begin{figure*}
 \centering\includegraphics[viewport=0 0 864 700, clip=true, width=\textwidth]{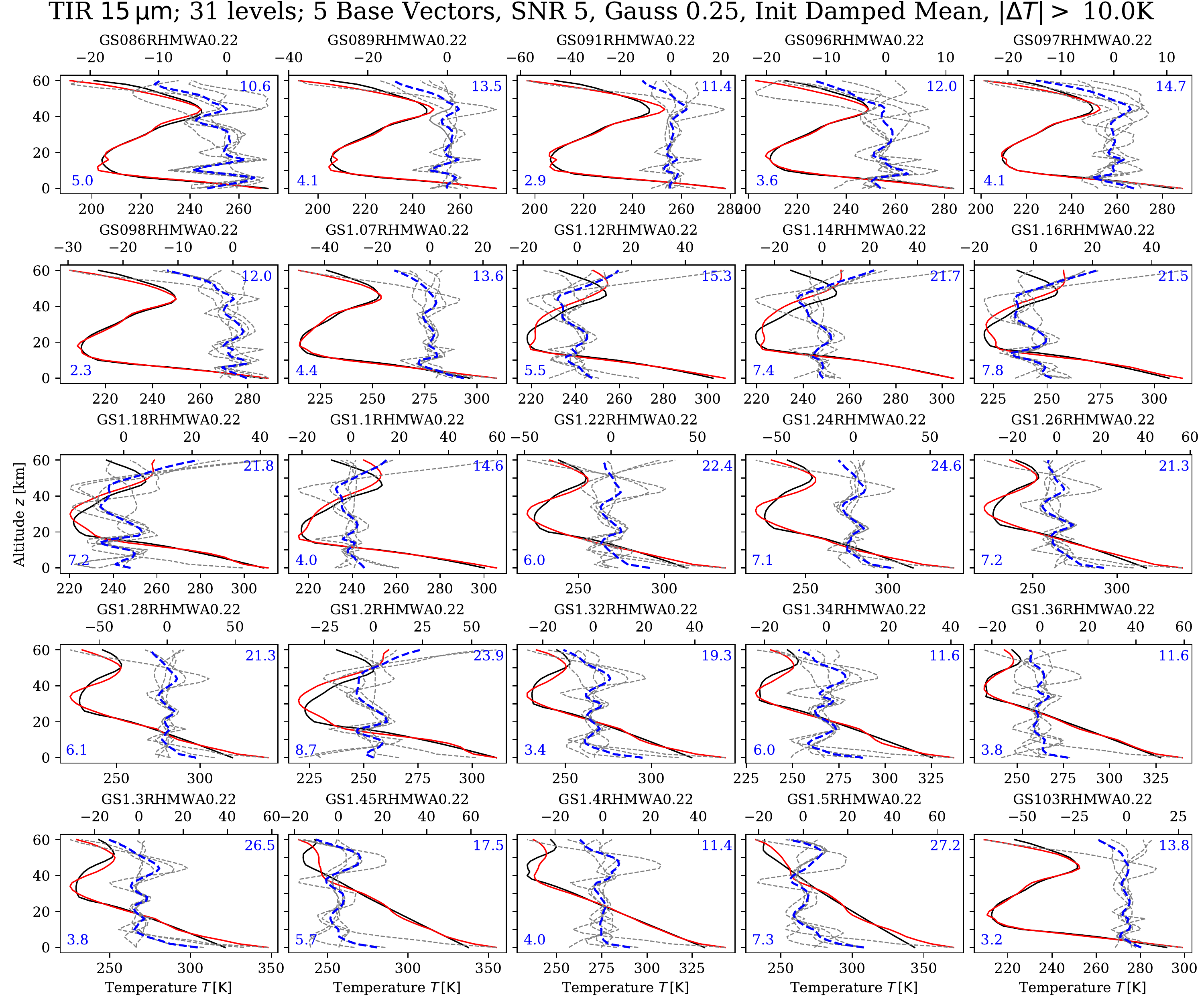}
 \caption{TIR-LW nonlinear least squares temperature fits of the G-star planets.
          Only ``poor'' fits with mean temperature differences (number in top right corner) larger than 10\,K are shown.
          The number in the lower left corner is the maximum temperature difference in the UTLS for this mean.
          Line colors and styles as before.
          The naming convention given in the title indicates the stellar insolation, the relative humidity according to \citet{Manabe67}, along with surface albedo 0.22.
          }
 \label{GPlanets}
\end{figure*}

Further retrievals have therefore been performed with the base vectors from the SVD of the Earth (Garand) matrix, the G-temperature matrix, the combined Earth and G-planets matrix, and for the matrix comprising all three sets of temperature profiles.
Figure \ref{GPlanets} summarizes the results of these fits in the LW TIR.
For some cases, differences in the UTLS are larger than $10 \rm\, K$ for the individual fits, but the error of the combined fit is always less than $9 \rm\, K$.

\begin{figure*}
 \centering\includegraphics[width=\linewidth]{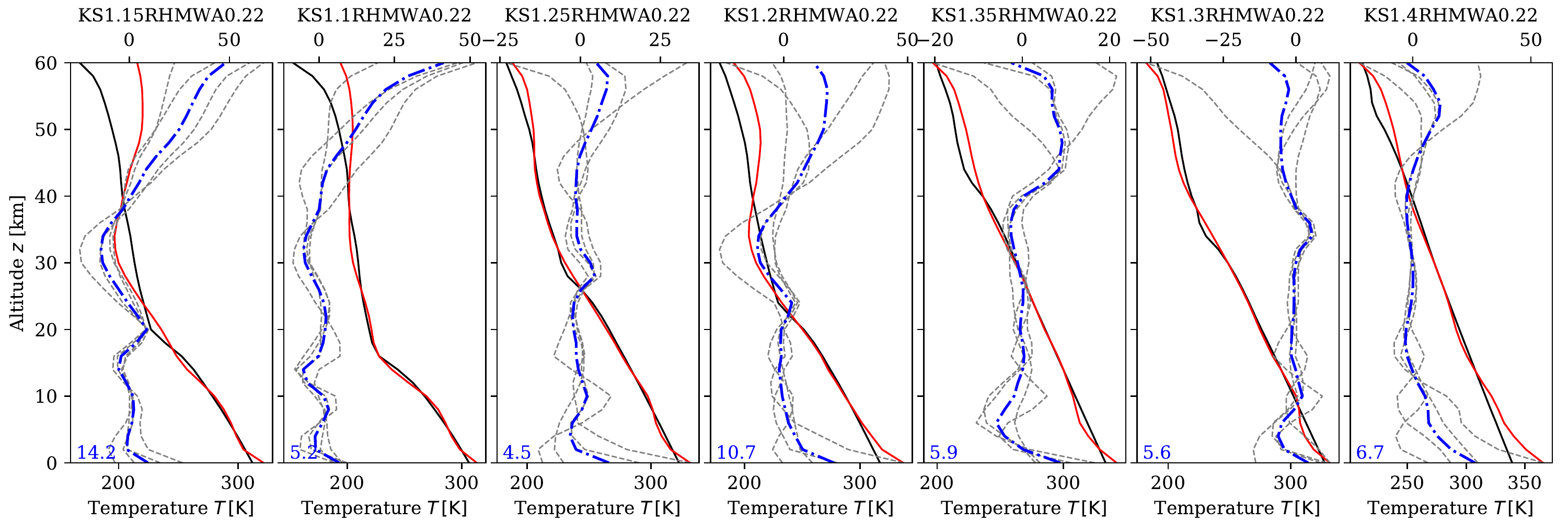}
 \caption{TIR nonlinear least squares temperature fits of the K-star planets. Line colors, styles, and planet naming as before.}
 \label{KPlanets}
\end{figure*}

Figure \ref{KPlanets} depicts the temperature retrieved from S/N=5 synthetic K-star planet observations.
Retrievals have been performed for both TIR SW and LW regions and for two sets of base vectors: one resulting from the combined Earth (Garand) and M-Earth temperature matrix and another temperature matrix also including the K-planet temperatures.
Apart from two cases the mean of these fits has deviations in the UTLS less than $10\rm\,K$.

\section{Discussion}
\label{sec:disc}

\subsection{Model assumptions and approximations}
\label{ssec:disc1}

The main objective of this feasibility study has been an assessment of the discretization by function expansion for the temperature retrieval; computational speed has not been a concern.
Accordingly, we used Py4CAtS \citep{Schreier19p} with approximate derivatives as opposed to GARLIC \citep{Schreier14} which computes exact derivatives.
Algorithmic (or analytic) differentiation (AD) can deliver significantly faster code:
a speed-up in code execution by a factor of twelve has been reported in \citet{Schreier15} for AD Jacobians compared to finite difference Jacobians.
Moreover, finite difference derivatives can exhibit substantial errors.
Nevertheless, the quality of almost all fits presented here indicates that the finite difference approximation used by MINPACK is working well.

For the cases considered, a range of approximations were used for the retrievals shown so far.
First, the surface emission term in the Schwarzschild equation \eqref{schwarzschild} has been neglected.
This does not have a large impact on the radiance in the center of the band with strong atmospheric absorption but will affect the band wings.
For an assessment of its impact, further retrievals were conducted, which suggested an negligible effect on the retrieved temperature.
However, for an analysis of real observations, this term has to be included in the forward modeling, but for the analysis of synthetic measurements, the consistency of the model assumptions is important.

Furthermore, it should be noted that a nadir geometry with a viewing angle of $180^\circ$ from zenith has been assumed for all runs.
For real measurements a representative viewing angle of about $144^\circ$ from zenith is more appropriate (or an integral or sum accounting for the disk averaging)
but for the synthetic measurements used here, the consistency of model assumptions is more important than the actual angle.
Retrieval runs with a $144^\circ$ angle essentially confirm the $180^\circ$ runs.

Moreover, we assumed an instrument spectral response function characterized by a constant HWHM $\Gamma$ whereas for many instruments (e.g.,\ gratings) a constant resolution $R = \nu / \delta\nu$
would be more appropriate.
In particular, MIRI MRS offers resolutions $R=1500 \text{--} 3500$ and NIRSpec has $R=1000$ and $R=2700$ \citep{Stevenson16}.
However, for the spectral intervals considered here (i.e., the TIR regions $2350 \text{--} 2420\cm$ and $660 \text{--} 720\cm$) the differences are only minor.

In order to speed up the fits, only the main absorbing gases have been considered in addition to \element{CO_2} (or \element{O_2} in the microwave), that is,\ \element{O_3} (in the LW TIR), \element{CH_4} (SW TIR), and
\element{H_2O} (see \qufig{optDepth} for a comparison of molecular optical depths).
For Earth's midlatitude summer atmosphere nitrous oxide changes the equivalent brightness temperature spectrum in the SW-TIR by less than a hundredth of a Kelvin.
Likewise, methane changes the temperature spectrum by some hundredths of a Kelvin.
Hence, both trace gases have a negligible impact on the total brightness spectrum of Earth.
However, \element{CH_4} is strongly enhanced in some of the M-Earth atmospheres, and the concentration of \element{N_2O} is larger especially in the upper atmosphere of some M-Earths.
In view of the modest concentration increase in the lower atmosphere and the negligible contribution to Earth's radiation, \element{N_2O} is not expected to have a significant influence upon the TIR-SW retrievals.

As a further simplification we used the HITRAN 1986 version \citep{Rothman87} for all simulations
(3\,353 lines of \element{H_2O}, \element{CO_2}, \element{N_2O} and \element{CH_4} for $2330 \text{--} 2440 \rm\,cm^{-1}$ and
15\,861 lines of \element{H_2O}, \element{CO_2}, \element{O_3} and \element{N_2O}  for $650 \text{--} 720 \rm\,cm^{-1}$),
whereas the most recent 2016 version \citep{Gordon17etal} comprises 24\,578 lines in the TIR-SW (with \element{N_2O} additionally) and 82\,595 lines in the TIR-LW (with \element{CH_4} additionally).
In the LW TIR, the equivalent brightness temperature changes less than 0.2\,K for small wavenumbers ($\nu < 690\cm$, probing the upper atmosphere, cf.\ \qufig{wgtFct_tir2}), but differences up to almost 2\,K are visible at large wavenumbers in the wing of the \element{CO_2} band.
In the SW TIR differences of the equivalent brightness temperature up to 1.15\,K (at $\nu \approx 2395\cm$) are observed.
Obviously the accuracy and completeness of the molecular spectroscopic database is important for analysis of real observations.

For the setup of the matrix to be decomposed, a common altitude grid was mandatory for all temperatures.
The available data were often confined to altitudes slightly above 60\,km and, hence, all data were regridded to a uniform grid with ToA at 60\,km.
However, thermal radiation originating from upper atmospheric layers is likely to impact the observed upwelling radiance.
Simulations indicate that the equivalent brightness temperature will change up to $3\rm\,K$ in the LW TIR and by less than $1.5\rm\,K$ in the SW (see \qufig{mesosphere}).
Hence, the upper atmosphere should be included in the forward model for accurate retrievals.
Extrapolation of an atmospheric data set to higher altitudes is probably straightforward for pressure, but an erroneous extrapolation of the temperature profile might deteriorate the fit.
For retrievals of actual observations, the provision of atmospheric data with sufficient altitude coverage is, therefore, important.

For many observations, the presence of aerosols and clouds may also prove to be an issue \citep[e.g.,][]{Kreidberg14} but these have been neglected in our radiative transfer (Schwarzschild equation \eqref{schwarzschild}).
To a first approximation, clouds will mask the lower atmosphere and, thereby, the information about temperature.
The weighting functions shown by \citet{Vasquez13c} demonstrate that contributions from below low-level water clouds will be blocked almost completely.
\citet{Kitzmann11r} discussed the impact of clouds on the retrievability of surface temperatures from low-resolution IR spectra.

For all retrievals, we used either the shortwave TIR with the \element{CO_2} $\nu_3$ band around 4.3\mue\ or the longwave TIR with the $\nu_2$ band around 15\mue, both of which  are covered by the JWST NIRSpec and MIRI instruments, respectively.
The combination (i.e.,\ mean) of the two fitted temperature profiles can be used to further improve the estimate.
For meteorological temperature soundings, further bands (channels) are exploited, in particular the $2150 \text{--} 2250 \cm$ ($4.44 \text{--} 4.65\mue$) and $1210 \text{--} 1650 \cm$ ($6 \text{--} 8.26\mue$) regions.
Both intervals can also be observed by the two JWST instruments to further improve the temperature retrievals.

\begin{figure*}
 \centering\includegraphics[width=\textwidth]{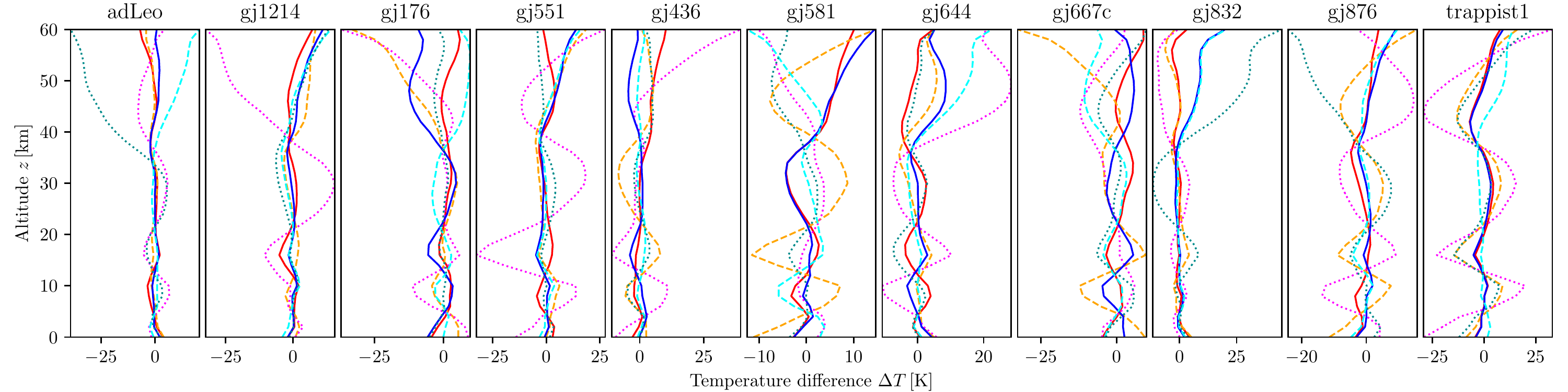}
 \caption{Errors of M-Earth temperature retrievals with reduced resolution.
          $\rm S/N=5$. Reddish colors: TIR-LW; bluish colors: TIR-SW;
          solid lines: high resolution ($R=2800$ and $R=2400$ for LW and SW, respectively); dashed lines: resolution halved; dotted lines: resolution quartered.}
 \label{MEarth-resolution}
\end{figure*}

\subsection{Atmospheric composition}
\label{ssec:disc2}

The M-Earth retrievals demonstrate that the fits are relatively insensitive to the precise concentration of interfering trace gases.
Small variations of the \element{CO_2} profile also have little impact on the retrieval.
However, a significantly reduced \element{CO_2} mixing ratio leads to strongly oscillating temperature profiles, whereas \element{CO_2} enhanced to 1000\,ppm or more tends to overestimated tropospheric
temperatures or zigzag stratospheric temperatures.

Hence, concurrent fitting of the \element{CO_2} concentration would clearly be desirable.
However, adding \element{CO_2} to the state vector of unknowns significantly deteriorates the condition of the Jacobian matrix (see subsection \ref{ssec:svd}), which would make the least squares fit much
more challenging due to degeneracy.

Knowledge of the pressure profile is expected to be more important and, fortunately, the atmospheric scale height can be estimated from transit spectroscopy.
For example, \citet{Paris13r} found that surface conditions (pressure and temperature) of terrestrial planets can be inferred from TIR emission spectra under idealised conditions.
Admittedly, transit and emission spectroscopy probe different regions of the atmosphere;
however, because of its long lifetime, \element{CO_2} can be expected to be uniformly distributed and an estimate of its VMR at the terminator should be useful.

For a further assessment, fits of the M-Earths were performed with the pressure scaled in the forward model data.
For the SW TIR, some of the fits with halved pressure were terminated because negative temperatures were obtained during the iteration.
All fits with double and fourfold pressure converged, however, for more than 50\% of the cases the largest UTLS temperature error was larger than 10\,K (worst case 20 and 32\,K for GJ\,832).
The LW TIR fits were slightly better: all fits converged, fits with doubled or quadrupled pressure prove somewhat better than fits with reduced pressure, and GJ\,832 is usually the most difficult case.
Considering that almost all fits were successful (i.e., all converged and delivered reasonable temperatures) when using as input the correct model pressure,
the failure (termination) of a fit can be interpreted as a hint that an inappropriate pressure (or more generally, atmospheric model) was applied.

In view of the difficulties involved when retrieving atmospheric temperature and composition simultaneously from emission spectra, observation and analysis of transit spectra becomes even more important.
Alternatively, estimates of \element{CO_2} partial pressures could be provided by coupled interior-atmosphere models \citep[e.g.,][]{Noack14,Tosi17,Godolt19}.
Data could also be provided by appropriate climate-chemistry modeling, as in \citet{Segura03,Rauer11,Grenfell14,Scheucher18,Wunderlich19}.

\subsection{Resolution, noise, and number of eclipses}
\label{ssec:disc3}

The S/N=5 adopted here for most retrieval tests is extremely bad from an Earth remote sensing point of view but still very optimistic for observations of Earth-like exoplanets.
Furthermore, we had assumed the high resolution modes of JWST MIRI and NIRSpec because the weighting function analysis (\qufig{wgtFct_tir2} and \ref{wgtFct_tir3}) indicates that high resolution is needed to retain the sensitivity to the upper atmosphere (see also the microwave retrieval tests of Subsection \ref{ssec:garand1}).
Retrieval tests with reduced resolution confirm this expectation:
whereas the fits of the M-Earth temperatures with halved resolution ($R=1400$ and $R=1200$ in the LW and SW TIR, respectively) are still acceptable (with maximum UTLS deviations larger than $10\rm\,K$ for three LW
fits only), a further reduction of the resolution by another factor two significantly deteriorates the estimated temperatures (see \qufig{MEarth-resolution}).

\begin{figure}
 \centering\includegraphics[width=\linewidth]{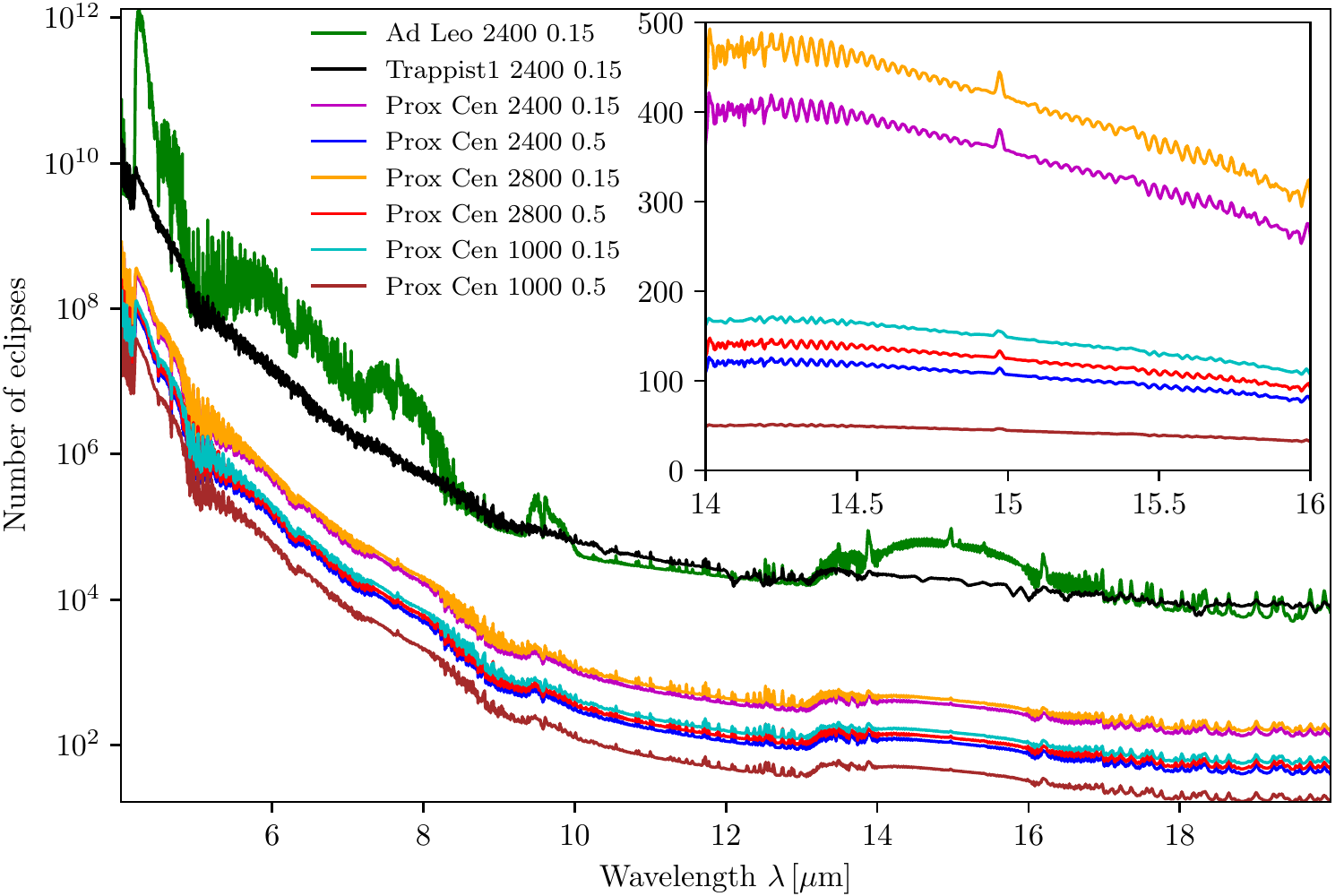}
 \caption{Number of eclipses required for S/N=5 observations of ``Earth'' at Proxima Centauri (distance 1.3\,pc), AD Leo (4.9\,pc), and Trappist 1 (12.1\,pc).
          The numbers in the legend indicate the resolution $R$ and the throughput.}
 \label{nTransits}
\end{figure}

Figure \ref{nTransits} depicts the number of eclipses required for an S/N=5 observation of some hypothetical Earth-like planets around M-dwarfs.
The values represent the required observing time for an JWST-like instrument (telescope detector area $25.4\rm\,m^2$ \citep{Kalirai18}) with a resolution and throughput (15\%) constant over wavelength.
We account for photon noise only to give an upper limit on the detection feasibility \citep{Rauer11}; saturation effects, readout noise, thermal noise, dark noise, and zodiacal noise are not considered
\citep[see also][]{Wunderlich19}.

For all hypothetical Earth-like planets a large number of secondary eclipses is required to reach an S/N of 5.
The number of eclipses is approximately proportional to the resolution and proportional to the square root of S/N.
The LW TIR is clearly advantageous because of the higher planet-star contrast compared to shorter wavelengths.
For the best case, an Earth-like planet around Proxima Centauri, an S/N of 2 could be obtained with a more feasible number of 7 eclipses (also see \qufig{MEarth-snr}).
The results are consistent with other studies concluding that Earth-like planets are probably too faint to reach high S/N with JWST emission spectroscopy, even with much lower resolution
\citep[see e.g.,][]{Rauer11,LustigYaeger19}.

The apparent difficulties of inferring the state of terrestrial exoplanetary atmospheres can be considered as an intrinsic problem that is also faced by other inversion techniques.
Fortunately, future missions such as the Origins Space Telescope (OST) will greatly enhance infrared remote sensing of Earth-like exoplanets.
OST \citep{Battersby19} is the only infrared instrument among several large mission concepts being considered by
NASA, aiming for increased sensitivity of more than a factor of a thousand  compared to prior missions.
(HabEx \citep{Gaudi19} and LUVOIR \citep{Roberge19} will also observe exoplanets, but with limitations to the UV and near IR.)
Alternatively LIFE, the Large Interferometer for Exoplanets, is currently designed for the characterization of terrestrial exoplanet atmospheres \citep{Defrere18,Quanz19}.

\section{Conclusions}
\label{sec:concl}

Temperature profiles of Earth-like exoplanets orbiting G-, K-, and M-stars were retrieved from synthetic thermal infrared emission spectra using a standard nonlinear least squares solver coupled with a line-by-line radiative transfer code.
The profiles have been represented using the leading singular vectors resulting from the SVD of a matrix comprising a representative set of temperatures
and the expansion coefficients are considered as the unknowns of the fit (i.e., the state vector).

Tests with synthetic Earth observations modeled with the ``\citet{Garand01} climatology'' revealed that even for $\text{S/N}=5$ the temperature can be retrieved with errors (deviations to the true profile) of less than $10 \rm\, K$ in the UTLS ($6 \text{--} 40\rm\,km$).
Larger errors can show up in the upper atmosphere.
The LW TIR retrievals are mostly better than the SW TIR retrievals (although the latter could be improved with an extended spectral range).

The analysis of synthetic M-Earth observations (again with $\text{S/N}=5$) also delivered temperatures largely consistent in the mid atmosphere (both in the SW and LW TIR) and relatively insensitive to the atmospheric model data used by the iterative least squares solver.
Likewise, retrievals of the G- and K-star planets were quite reasonable with, however, larger deviations confined to the ToA.
In all cases (Earth and exoplanets), moderate resolution (about 2\,500, possible with JWST) is crucial to retain sensitivity to the upper atmosphere.

Detection of a temperature inversion in exoplanetary atmospheres is particularly valuable \citep[e.g.,][]{Crossfield15,Madhusudhan19,Malik19}.
With a few exceptions, temperatures in the upper troposphere and lower stratosphere were retrieved with deviations from the true profiles mostly smaller than 10\,K.
For the ``failed fits'' (i.e.,\ Earth atmospheres \#16 and \#31 in SW-TIR) the estimated tropopause minimum was substantially too small
(however, the fits were much better with an enlarged fitting window) but, nevertheless, the shape was correct and the inversion was detected.
The retrievals of the G-, K-, and M-star planetary emission spectra also revealed the shape of the profile correctly.

With five unknowns (see also \qufig{mWave_nBase}), our inverse problem is comparable in size with the retrievals utilizing temperature profile parameterizations \citep[e.g.,][]{Madhusudhan09}.
Clearly, these parameterizations are motivated by atmospheric physics, which can be both an advantage and disadvantage (i.e., biased on our solar system experience).
Our function expansion based on the SVD of sets of representative profiles also depends somewhat on a priori knowledge,
but expansions using other sets of base functions \citep[e.g.,][]{Doicu05b} would serve as an alternative.
Moreover, because of the generic nature of function expansion techniques and least squares, our approach can be readily adapted to the analysis of other classes of planets, such as super-Earths, mini-Neptunes, and hot-Jupiters.

For the analysis of future spectroscopic observations, several recommendations arise from our study: 
several fits should be performed with different base vectors (or number of base vectors) and different initial guesses.
Furthermore, different model atmospheres (i.e.,\ pressure and gas concentration profiles, esp.\ for \element{CO_2}) should be considered.
Finally, (slightly) different spectral windows can be tested (assuming the observed spectrum is large enough) and fits with (slightly) different altitude grids for the model atmospheric data might also give some clues for the quality of retrievals.
Accompanying transit spectroscopy and atmospheric modeling would be highly valuable in future studies.


\begin{acknowledgements}
This research was supported by DFG projects SCHR 1125/3-1, RA-714/7-1, and GO 2610/1-1.
Furthermore we acknowledge the support of the DFG priority programme SPP 1992 "Exploring the Diversity of Extrasolar Planets (GO 2610/2-1)".
We would also like to thank Adrian Doicu and Thomas Trautmann (Oberpfaffenhofen) and Heike Rauer (Berlin) for useful discussions and critical reading of the manuscript.
\end{acknowledgements}


\begin{appendix}

\section{Data}

\begin{figure}
 \centering
\includegraphics[width=\linewidth]{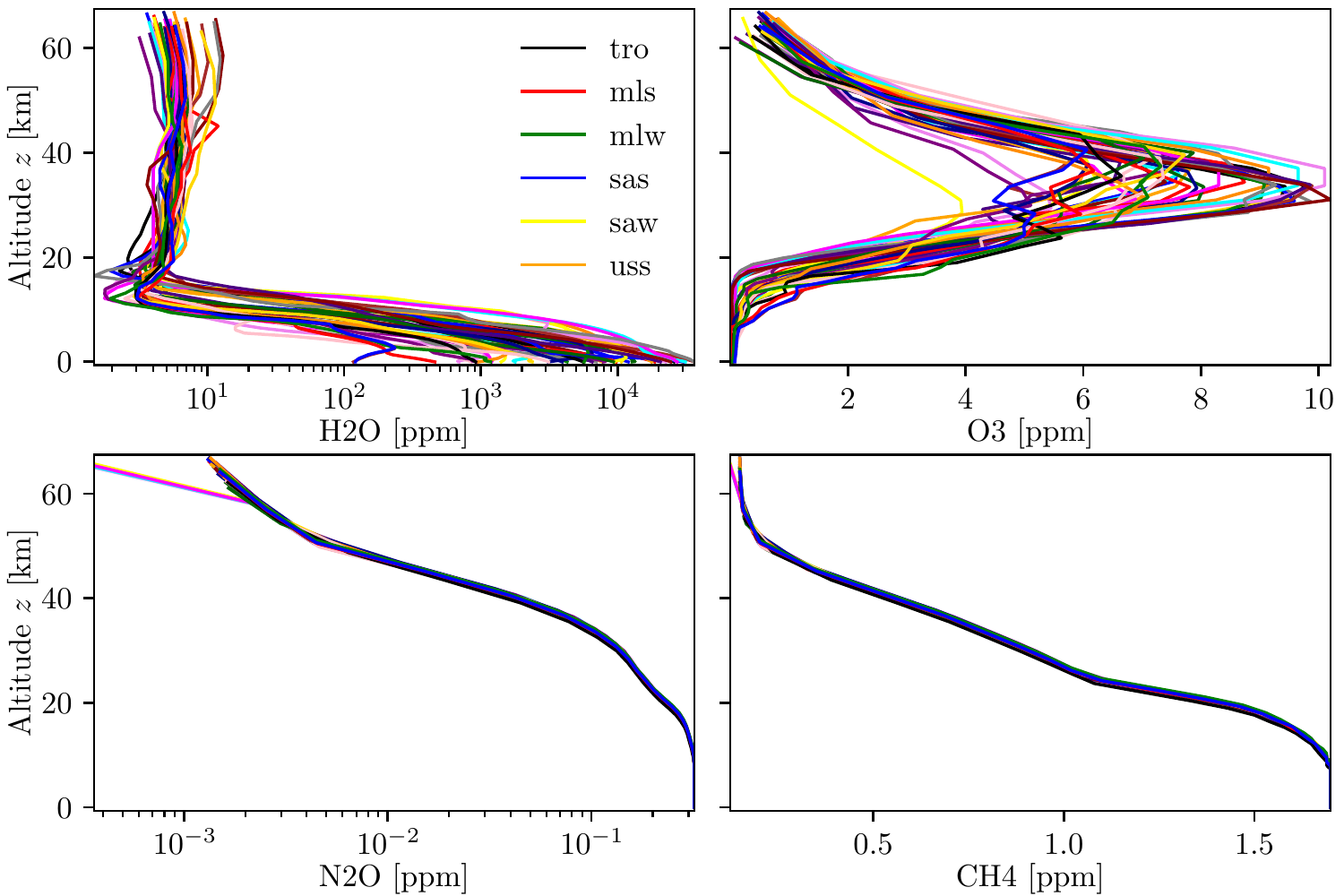}
 \caption{Volume mixing ratios of water vapor \element{H_2O}, ozone \element{O_3}, nitrous oxide \element{N_2O}, and methane \element{CH_4} profiles of the 42 Garand atmospheres.}
 \label{garand_gases} 
\end{figure}

\begin{figure}
 \centering
\includegraphics[width=\linewidth]{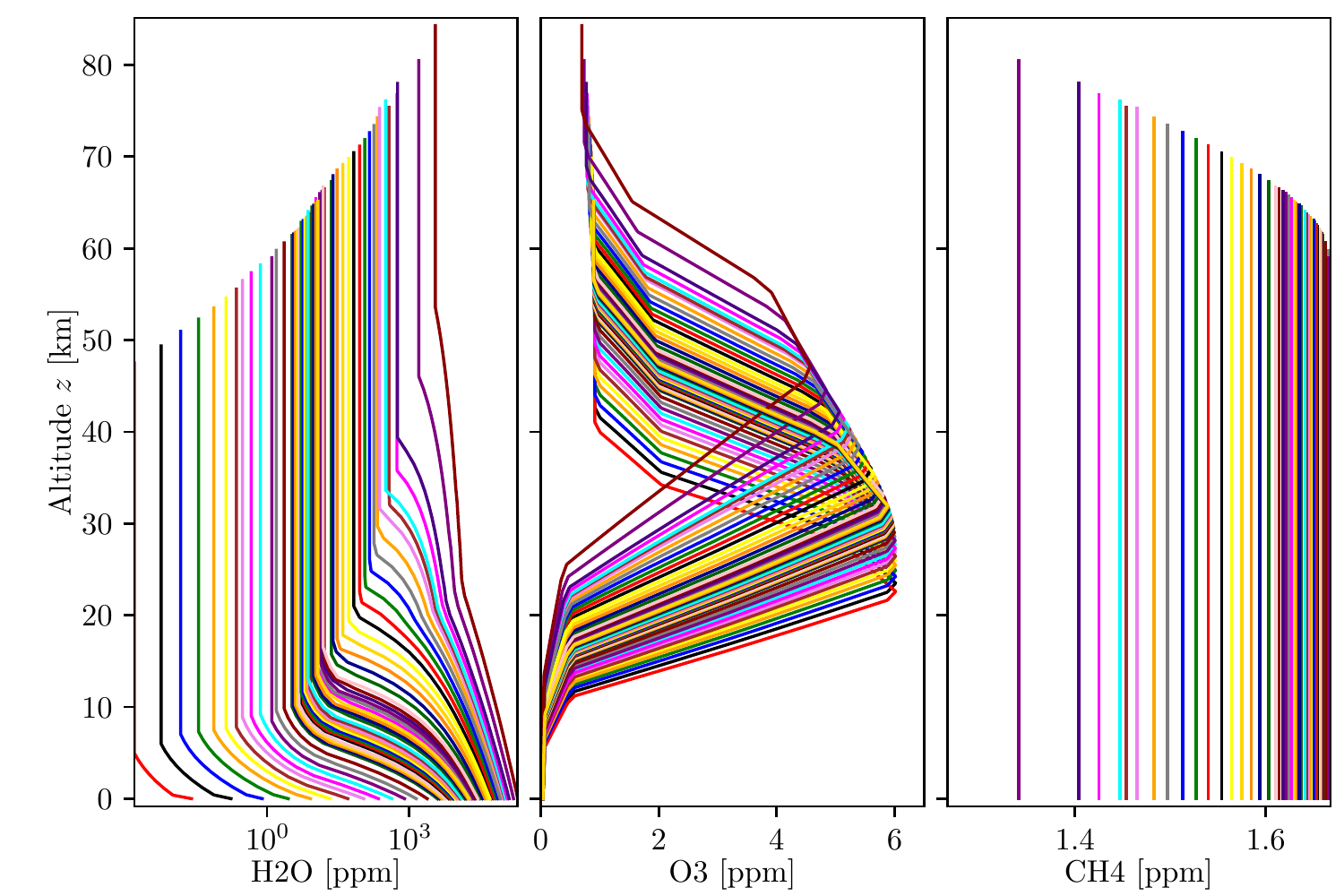} 

\includegraphics[width=\linewidth]{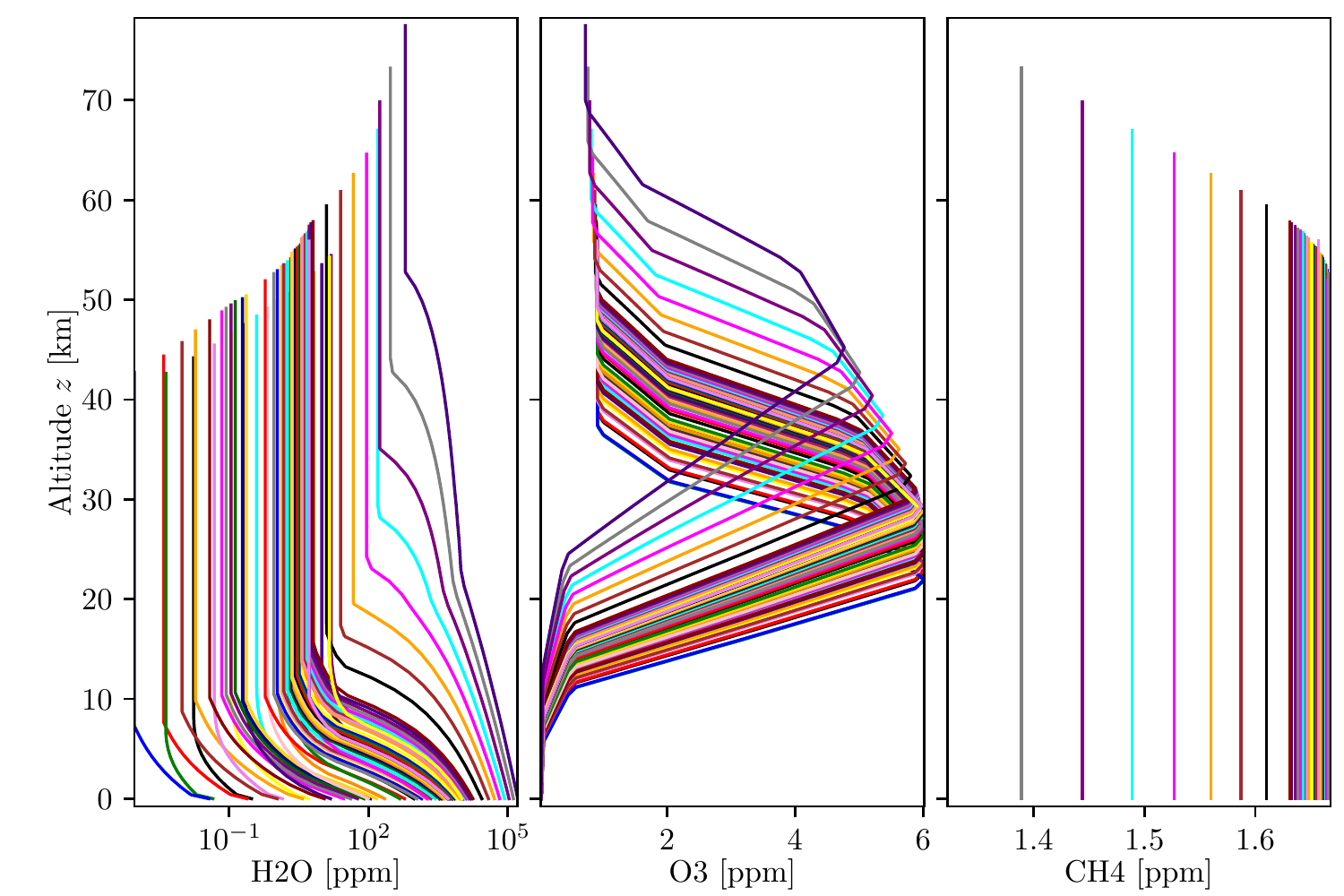} 

\includegraphics[width=\linewidth]{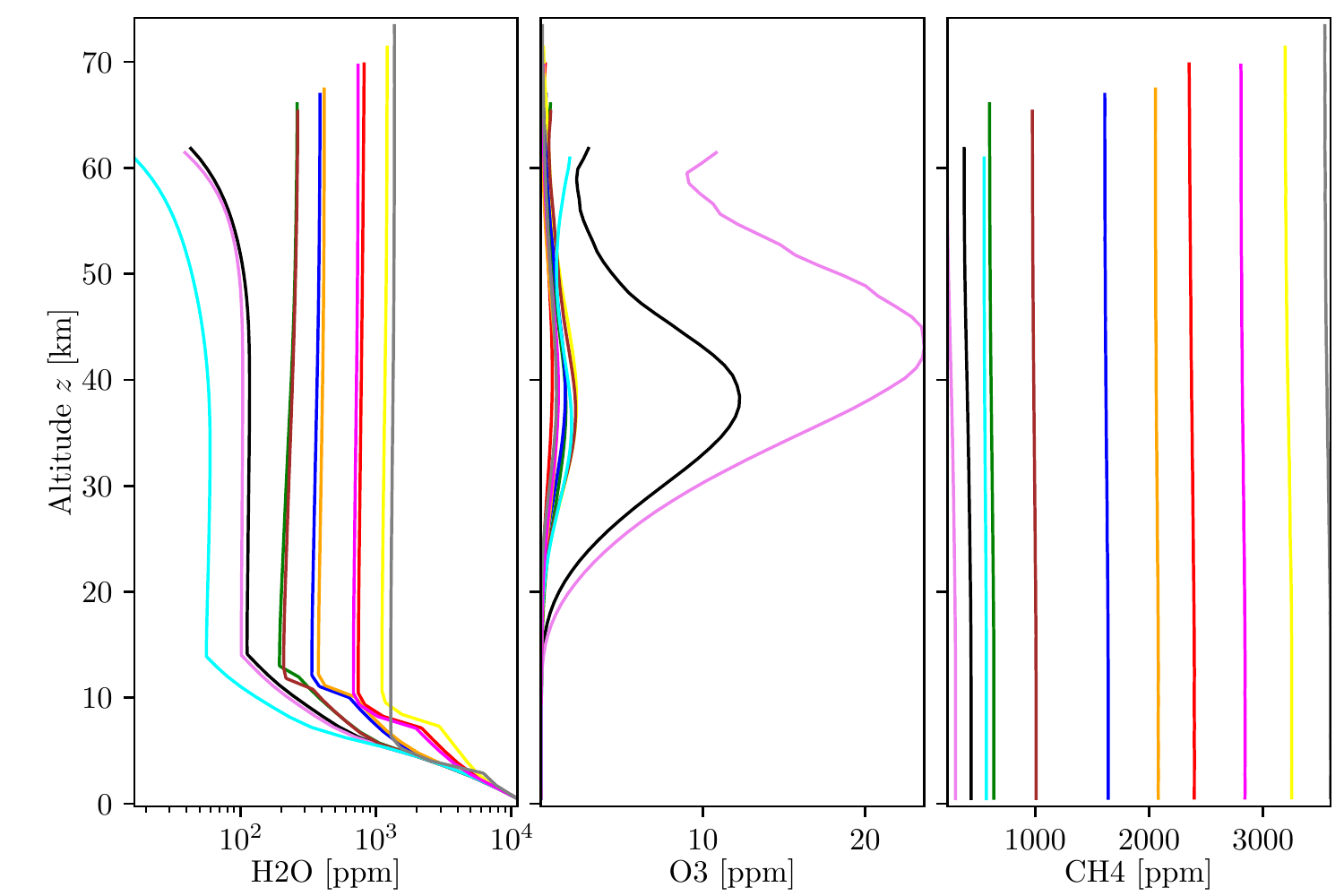}
 \caption{\element{H_2O}, \element{O_3}, and \element{CH_4} profiles of the G-star (top), K-star (middle), and M-dwarf (bottom) planets. (Note that the M dwarfs data set includes ``Earth around Sun''.)}
 \label{gkmData} 
\end{figure}

\begin{figure*}
 \centering
\includegraphics[width=0.4\linewidth]{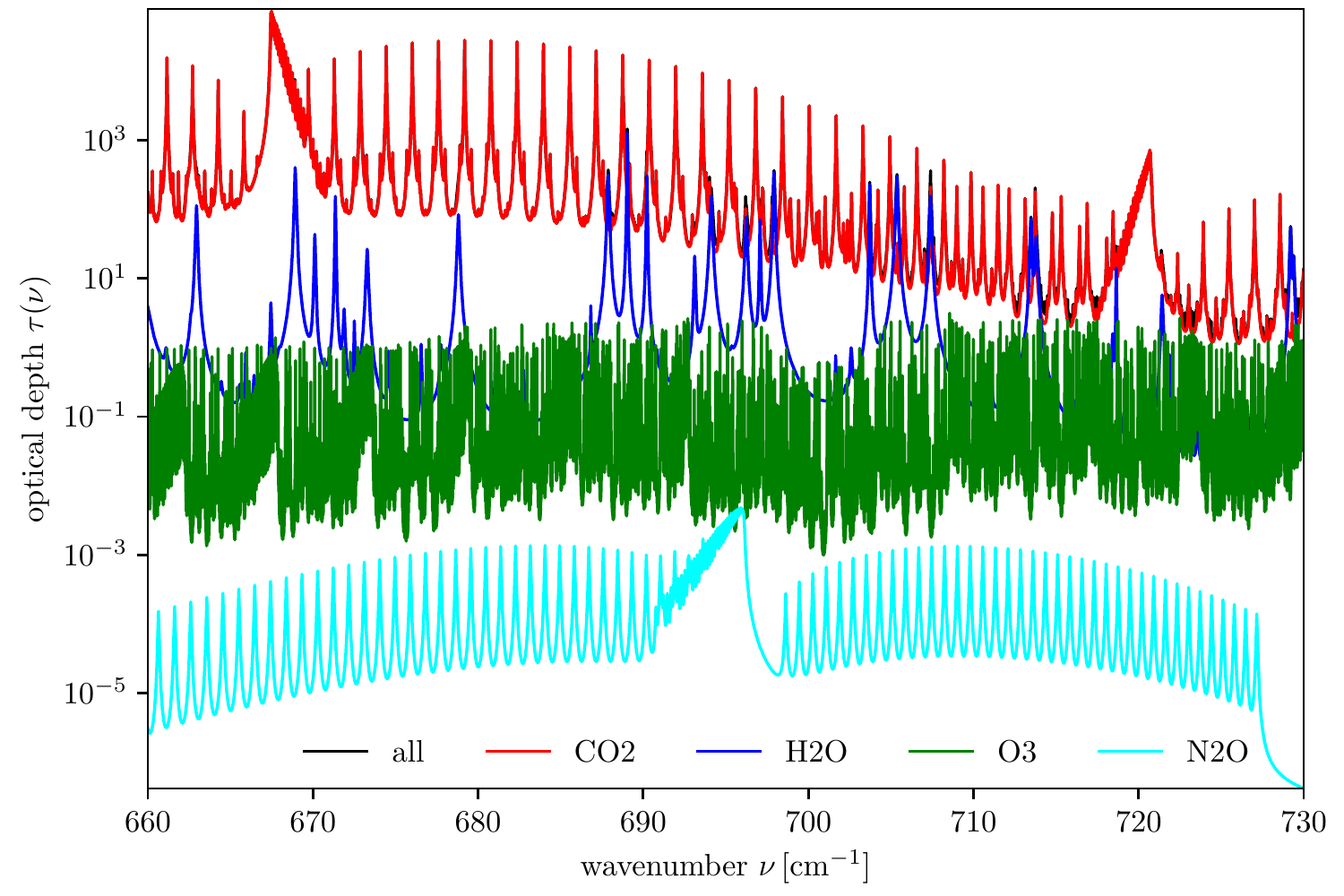}
 \includegraphics[width=0.4\linewidth]{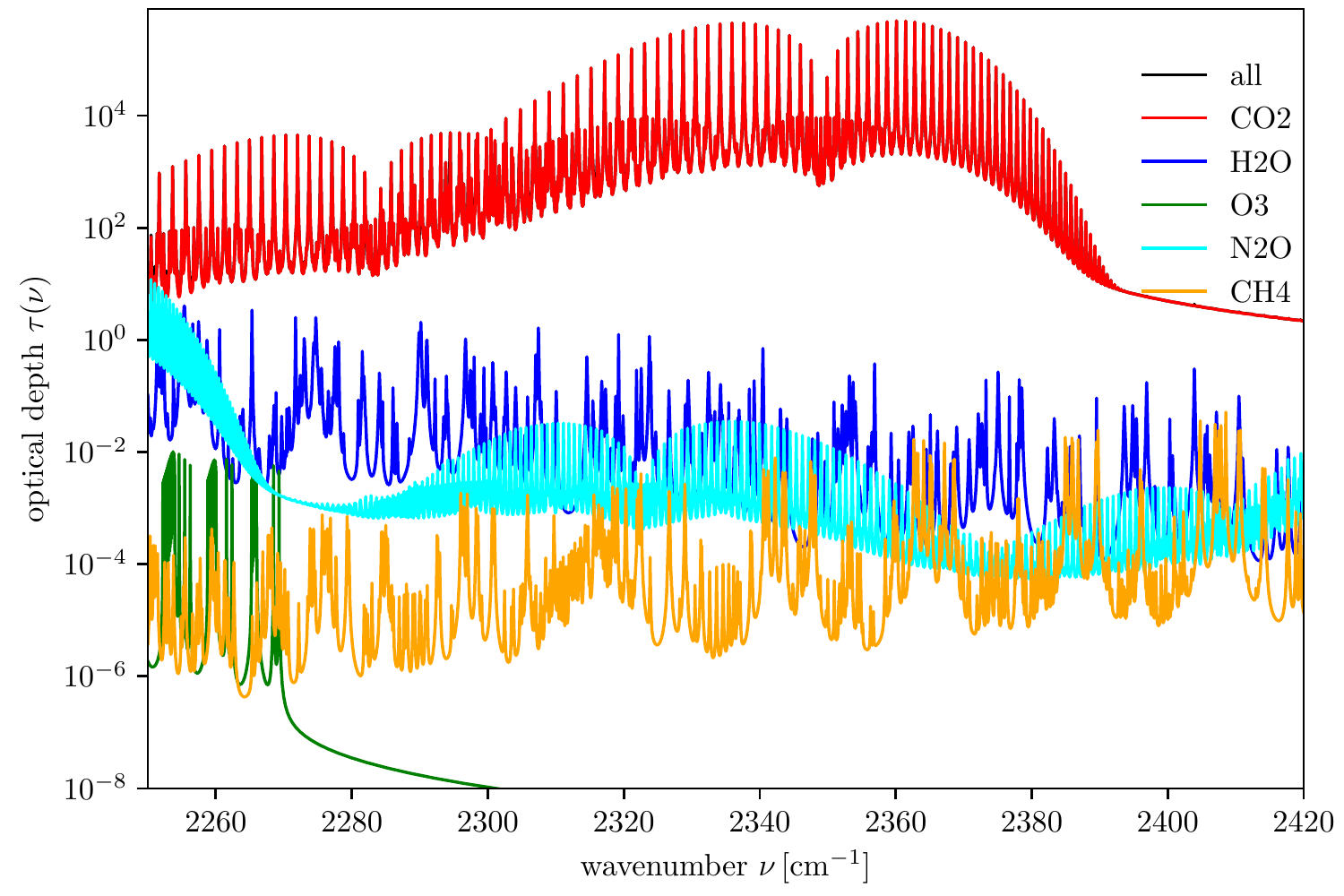}
 \caption{Comparison of molecular optical depths (based on HITRAN86).}
 \label{optDepth} 
\end{figure*}

\begin{figure*}
 \centering
\includegraphics[width=0.85\textwidth]{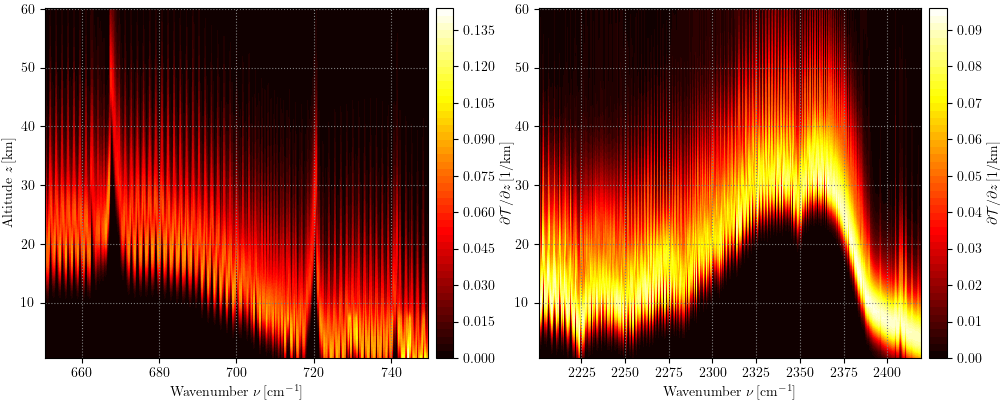} 
 \caption{Weighting functions for Earth around Proxima Centauri.}
 \label{proxCenWgtFct}
\end{figure*}

\begin{figure*}
 \centering
\includegraphics[width=0.75\textwidth]{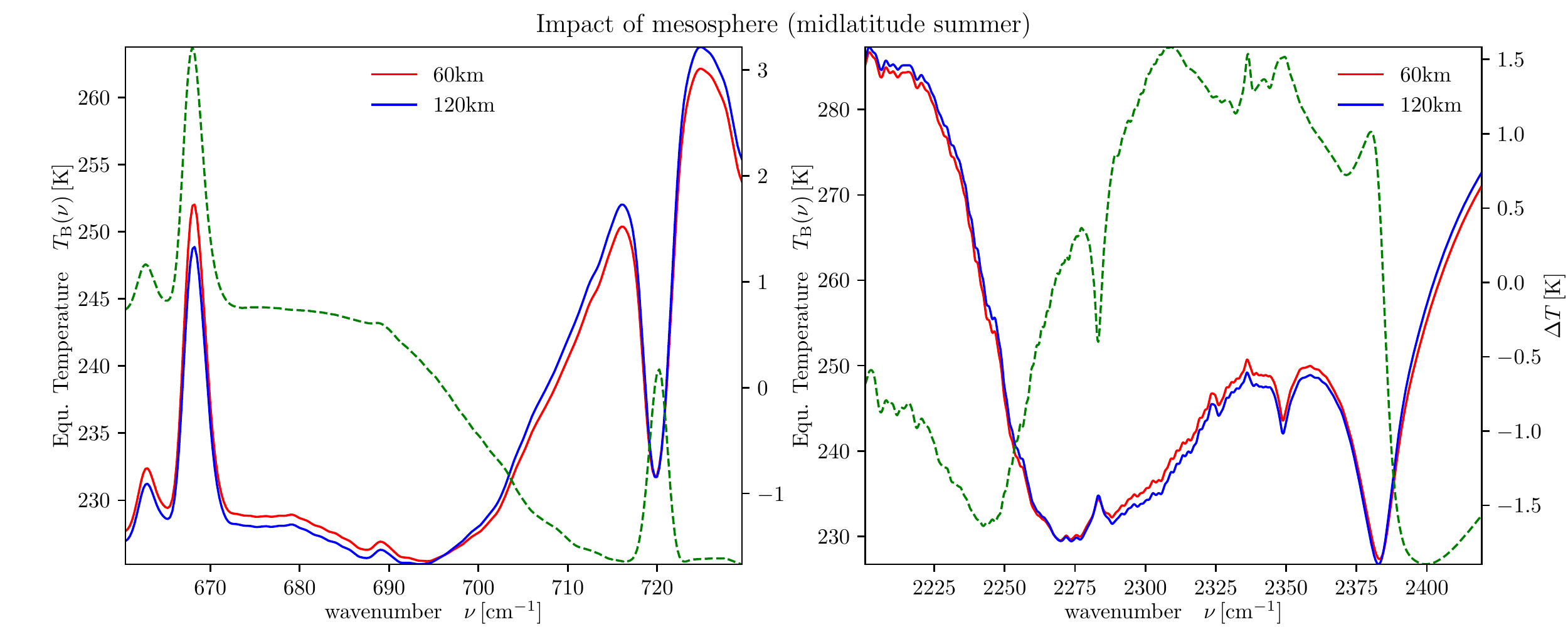}
 \caption{Impact of Earth's mesosphere on ToA radiance. Midlatitude summer atmosphere, Gaussian spectral response function.}
 \label{mesosphere} 
\end{figure*}

\begin{figure*}
 \centering
\includegraphics[width=0.75\textwidth]{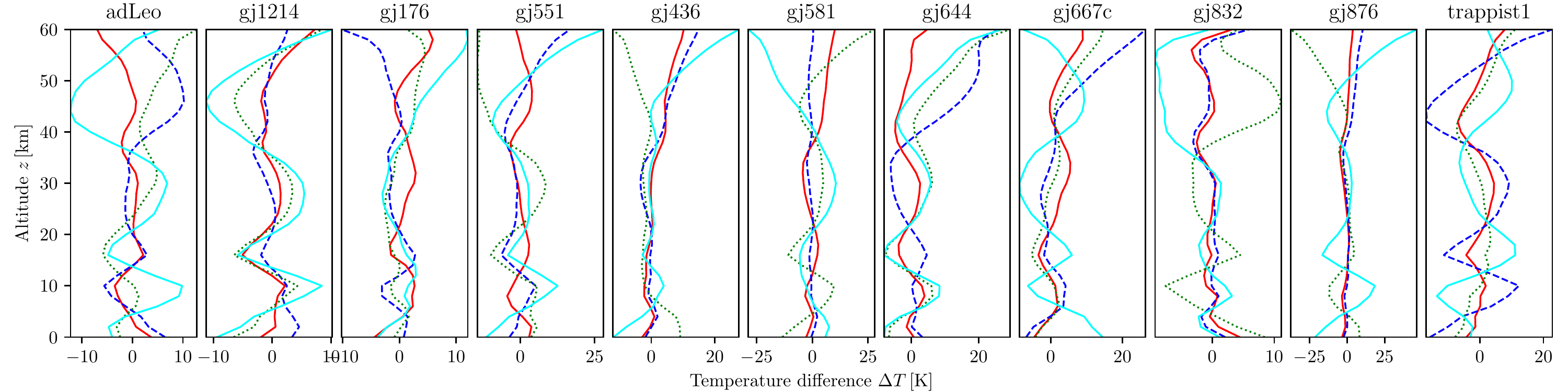}
 \caption{Errors of M-Earth temperature retrievals with increased noise. TIR-LW with $R=2800$.
          Red solid: S/N=5, blue dashed: S/N=4, green dash-dotted: S/N=3; cyan dotted: S/N=2.}
 \label{MEarth-snr} 
\end{figure*}

The objective of this online\emph{} appendix is to provide some supplementary material that might be useful for the reader.
Figures \ref{garand_gases} and \ref{gkmData} display the molecular concentration profiles.
The optical depths shown in \qufig{optDepth} clearly indicate the dominance of \element{CO_2} and the minimal interference of other species: in the band center the total optical depth and the \element{CO_2} optical depth are almost indistinguishable.
Figure \ref{proxCenWgtFct} demonstrates that the weighting function of our nearest neighbor exoplanet is similar to Earth's weighting function.
Figure \ref{mesosphere} indicates the relevance of Earth's mesospheric layers on the upwelling radiation.
Finally, the impact of even stronger noise on M-Earth retrievals is shown in \ref{MEarth-snr}.

\end{appendix}


\end{document}